\documentclass[twocolumn,numberedappendix]{emulateapj}
\bibpunct[; ]{(}{)}{;}{a}{}{;}
\usepackage{apjfonts}

\usepackage{graphicx} 
\usepackage{natbib}
\graphicspath{{../iraf/parallels/n602/}}  

\newcommand{\Msolar}{M${_\odot}$}

\shorttitle{Recent star formation in NGC\,602}
\shortauthors{De Marchi, Beccari \& Panagia}

\journalinfo{The Astrophysical Journal}
\received{11 April 2013}
\accepted{28 July 2013}

\slugcomment{}

\begin{document}

\title{Photometric determination of the mass accretion rates of pre-main
sequence stars. IV. \\ Recent star formation in NGC\,602\,\altaffilmark{*}}


\author{
Guido De Marchi,\altaffilmark{1}
Giacomo Beccari\altaffilmark{2}
and Nino Panagia,\altaffilmark{3,4,5}
}

\altaffiltext{1}{European Space Agency, 
Keplerlaan 1, 2200 AG Noordwijk, Netherlands; gdemarchi@rssd.esa.int}

\altaffiltext{2}{European Southern Observatory,
Karl--Schwarzschild-Str.2, 85748 Garching, Germany, gbeccari@eso.org}

\altaffiltext{3}{Space Telescope Science Institute, 3700 San Martin
Drive, Baltimore, MD 21218, USA, panagia@stsci.edu}

\altaffiltext{4}{INAF--NA, Osservatorio Astronomico di Capodimonte, 
Salita Moiariello, 16  80131 Naples, Italy}

\altaffiltext{5}{Supernova Limited, OYV \#131, Northsound Rd., Virgin
Gorda, British Virgin Islands, VG 1155}

\altaffiltext{{$\star$}}{Based on observations with the NASA/ESA
{\it Hubble Space Telescope}, obtained at the Space Telescope Science
Institute, which is operated by AURA, Inc., under NASA contract
NAS5-26555}

\begin{abstract}  

We have studied the young stellar populations in NGC\,602, in the Small
Magellanic Cloud, using a novel method that we have developed to
combine  {\em Hubble Space Telescope} photometry in the $V$, $I$, and
$H\alpha$ bands. We have identified about 300 pre-main sequence
(PMS) stars, all of which are still undergoing active mass accretion,
and have determined their physical parameters (effective temperature,
luminosity, age, mass and mass accretion rate). Our analysis shows that
star formation has been present in this field over the last
60\,Myr. In  addition, we can recognise at least two clear, distinct,
and prominent episodes in the recent past: one about 2\,Myr ago, but
still ongoing in  regions of higher nebulosity, and one (or more) older
than 30\,Myr, encompassing both stars dispersed in the field and two
smaller clusters located about $100\arcsec$ north of the centre of
NGC\,602. The relative locations of younger and older PMS stars do not
imply a causal effect or triggering of one generation on the other. The
strength of the two episodes appears to be comparable, but the
episode occurring more than 30\,Myr ago might have been even stronger
than the current one. We have investigated the evolution of  the mass
accretion rate $\dot M_{\rm acc}$ as a function of the stellar 
parameters finding that $\log \dot M_{\rm acc} \simeq -0.6\,\log t
+ \log m + c$, where $t$ is the age of the star, $m$ its mass and $c$ is
a decreasing function of the metallicity.

\end{abstract}


\section{Introduction}

The young star cluster NGC\,602 and the associated H\,II region N\,90
(Henize 1956) are located in a relatively isolated and diffuse
environment in the ``wing'' of the Small Magellanic Cloud (SMC), towards
the edge of the Magellanic Bridge. In many ways, NGC\,602 represents an 
extragalactic analogue to the greater Orion OB association: it has
a similar overall size  ($\sim 50$\,pc), it contains similarly few
O-type stars of about the same age ($\sim 2$\,Myr) and, like Orion, it
is located at the surface of a big molecular cloud in a ``blister''
opened by the radiation of the early stars. These young massive stars
and the associated H\,II region have been known and studied for almost
fifty years (Westerlund 1964; Hutchings et al. 1991; Battinelli \&
Demers 1992; Massey et al. 2000), but very little was known about
low-mass objects in the field. Recently, however, thanks to observations
made with the {\em Hubble  Space Telescope} (HST) and the {\em Spitzer
Space Telescope}, it has become possible to study the star formation in
NGC\,602 and surrounding regions over a wide range of masses and this
subject has been the topic of a number of recent papers.

Carlson et al. (2007) carried out a panchromatic study of NGC\,602 with
the HST at optical wavelengths and {\em Spitzer} in the range $3.6 -
8.0$\,$\mu$m (later extended to 24\,$\mu$m by Carlson et al. 2011),
discovering an extensive population of candidate pre-main sequence (PMS)
stars and young stellar objects (YSOs). Through comparison of the
observed magnitudes with theoretical PMS isochrones (although initially
not for the metallicity appropriate for the SMC) and with models of the
spectral energy distribution of YSOs, they conclude that star formation
started about 4\,Myr ago in the central cluster and later propagated
towards the periphery, where it still continues. A similar conclusion
was reached by Gouliermis et al. (2007), whose analysis of archival {\em
Spitzer} mid-infrared observations of an area of 30\,arcmin$^{2}$
around NGC\,602 reveals 22 YSO candidates, along the rims of the parent
molecular cloud. Like Carlson et al. (2007), Gouliermis et al. (2007)
attribute the formation of these YSOs to triggering caused by the
photoionisation of the young stellar association. 

Combining high resolution echelle spectroscopy and velocity maps from a 
neutral hydrogen (H\,I) survey (Stavely--Smith et al. 1997) with the HST
observations of Carlson et al. (2007), Nigra et al (2008) tried to
establish the likely scenario leading to the formation of NGC\,602.
These authors do not find morphological evidence of violent events such
as supernovae or strong stellar winds. They suggest instead that about
7\,Myr ago the interaction of two expanding H\,I shells created an
overdensity, from which NGC\,602 started to form about 3\,Myr later, and
newly formed massive stars started to erode the surrounding  nebular
material, creating a photodissociation region. These authors speculate
that NGC\,602 is the result of a single star-forming event, in a region
of low gas density. However, as we will show in this work, the star
formation history of this region is much more complex than what Nigra et
al. (2008) suggested. Indeed, Cignoni et al. (2009) have already shown
that, even though the current star formation episode at the centre of
NGC\,602 (hereafter NGC\,602\,A) had a peak about 2.5\,Myr ago, two
neighbouring associations appreciably older than it (NGC\,602\,B and
NGC\,602\,B2) were formed as part of the same process. According to
their analysis, these associations have ages between 15 and 150\,Myr
(Schmalzl et al. 2008 had previously suggested ages of 80 and 160\,Myr
for NGC\,602\,B and NGC\,602\,B2, respectively). 

To understand how star formation has proceeded in this field, it is
crucial to study the properties of PMS stars, particularly of those of
lower mass. These objects represent a living record of how star
formation has proceeded over the past $\sim 50$\,Myr or more, which is a
time span much longer than the one offered by the study of massive
stars. All the works cited above have indeed made use of PMS stars to
probe recent star formation. However, the selection of PMS objects in
those papers is based on a topological analysis in the broad-band
colour--magnitude diagram (CMD) alone, namely in the $V$ {\em vs.} $V-I$
plane. Therefore, those studies are limited to the youngest PMS stars, with
ages younger than $\sim 5$\,Myr, which are well separated in colour from
MS stars. In the CMD, these objects are located to the right of and
above the MS. Conversely, any PMS objects older than $\sim 10$\,Myr are
very hard to distinguish from the MS itself in the CMD, thereby
thwarting any attempt to study other recent star formation episodes in
the same field. 

On the other hand, thanks to distinctive emission features in the
spectra of PMS stars still undergoing mass accretion, we have shown that
it is possible to efficiently and reliably identify all objects of this
type in a stellar field, regardless of their age and of their position
in the CMD. Building on the work of Romaniello (1998), of Panagia et al.
(2000), and of Romaniello et al. (2004), De Marchi, Panagia \&
Romaniello (2010, hereafter Paper\,I) and De Marchi et al. (2011b,
hereafter Paper\,II) showed that through a suitable combination of
broad- and narrow-band photometry an accurate determination of the
H$\alpha$ luminosity of these objects is possible, from which the
accretion luminosity and mass accretion rate can be derived. In this
work, we apply the methods developed and tested in Papers\,I and II to
investigate how star formation has proceeded in the NGC\,602 region. 

The structure of the paper is as follows: in Section\,2 we describe the
observations and the photometric analysis, while we devote Section\,3 
to the search for PMS stars through their H$\alpha$ excess emission. In 
Section 4 we derive the physical parameters for these objects, including
effective temperature, luminosity, age, mass and accretion luminosity,
while Section\,5 addresses the presence of several generations of stars
in the field and their mutual relationships. In Section\,6 we derive the
mass accretion rate and  study how it evolves in time. A summary of the
most important conclusions of the paper is presented in Section\,7.

\section{Observations and data analysis}

This work is based on observations of a field of $\sim 3\farcm3 \times
3\farcm3$ around the cluster NGC\,602 obtained on 2004 July 14 and 18
with the Wide Field Channel of the Advanced Camera for Surveys (ACS) on
board of the HST (proposal nr. 10248, principal investigator A. Nota).
The dataset includes observations in the F555W band (hereafter $V$; five
exposures for a total duration of 2\,153\,s), F814W band (hereafter $I$;
five exposures for a total of 2\,265\,s) and N658N band (hereafter
$H\alpha$; three exposure for a total of 1\,908\,s). A suitable
dithering pattern and the combination of long and short exposures in the
same band provide proper sampling of the point spread function (PSF),
while preventing saturation of the brightest stars. 

\begin{figure}[t]
\centering
\resizebox{\hsize}{!}{\includegraphics[width=16cm,bb=0 0 504 350]{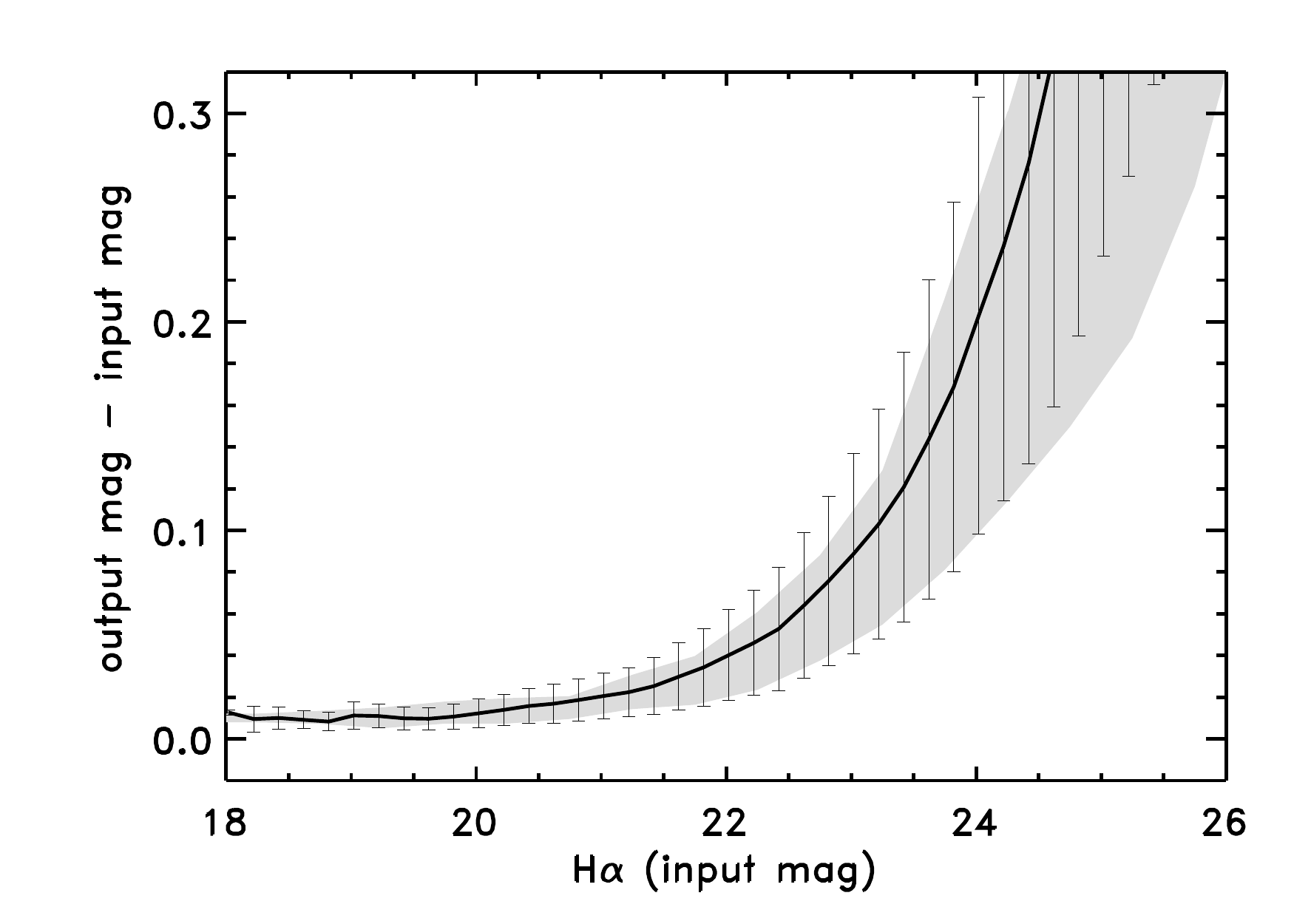}}
\caption{Mean of the error distribution of the $H\alpha$ magnitudes
(solid line), and associated $1\,\sigma$ uncertainties (error bars), 
obtained from the comparison between input and output magnitudes in the 
artificial star experiments. The shaded area shows the distribution of 
the photometric uncertainties calculated by ALLFRAME.} 
\label{fig1}
\end{figure}

As regards the photometry, we first analysed the $V$- and $I$-band
images, starting from the flat-fielded exposures corrected for geometric
distortion and for the effective area of each pixel (see Sirianni et al.
2005 for details). The PSF-fitting procedure DAOPHOT\,II (Stetson 1987)
was applied to each individual frame and a master list of stars was
obtained using all the stars detected in at least six of the ten $V$- and
$I$-band images. We derived in this way an initial catalogue free
of spurious detections, such as cosmic rays or haloes and spikes around
saturated stars, containing 7\,021 objects detected in the $V$ and $I$
bands, of which 6\,261 are also detected in the $H\alpha$ band. We
furthermore restricted the DAOPHOT\,II sharpness parameter $s$ to span
the range $-0.15\leq s \leq 0.15$, in order to exclude cosmic rays and
extended objects.

The resulting master list, containing 5\,788 stars, was then used
to perform PSF fitting over the entire dataset, including the $H\alpha$
images, using the standard ALLFRAME routine (Stetson 1994). The average
of the magnitudes measured in each individual frame for every master
list object was adopted as the star magnitude in the final catalogue,
while we took the resulting standard deviation around the mean for each
object as the associated photometric uncertainty (see e.g. Stetson
1987). Down to magnitudes $V=21.3$ and $I=21.0$ all stars detected in
$V$ and $I$ are also detected in $H\alpha$. This fraction drops
to 95\,\% at $V=28.5$ and $I=27.3$ and no stars detected in $H\alpha$
are fainter than these limits.

Since in the case of the $H\alpha$ images there are only three
exposures, we also conducted artificial star tests on those images in
order to independently verify the photometric uncertainty (see e.g.
Stetson \& Harris 1988). We added more than 170\,000 artificial stars,
spread over several tests and uniformly distributed with a density of
one object per $2.3$ square arcsec, in order to not alter the crowding
conditions of the frames. The input magnitudes of the artificial stars
were drawn from a luminosity function similar to the observed one but
monotonically extended well beyond the detection limit, in order to
increase the statistics for faint stars. We then applied the same
photometric procedure used for the original images and compared the
input magnitudes with the derived magnitudes, finding an excellent
agreement in their values, as shown in Figure\,\ref{fig1}. The
solid line corresponds to the mean difference and the error bars show
the $\pm 1\,\sigma$ uncertainties. In the analysis that follows we will
take these as photometric uncertainties in the $H\alpha$ band. In the
$V$ and $I$ bands the photometric uncertainties are considerably
smaller, owing to the wider pass bands. We also show for reference in
Figure\,\ref{fig1} the photometric uncertainty (shaded area) as derived
by ALLFRAME, which is slightly smaller than those derived through
artificial star experiments, as expected (see e.g. Beccari et al.
2013). In summary, the artificial star experiments gave us general
confidence that the uncertainties assigned by our photometry to the
magnitudes of individual stars is realistic, also in the $H\alpha$
band.

The final catalogue contains a total of 5\,500 stars, with well
defined magnitudes in the $V$, $I$ and $H\alpha$ bands, which were
calibrated into the VEGAMAG photometric system according to the
procedures described in Sirianni et al. (2005) and using the most
current photometric zero points (Bohlin 2012). 

\begin{figure}[t]
\centering
\resizebox{\hsize}{!}{\includegraphics[width=16cm]{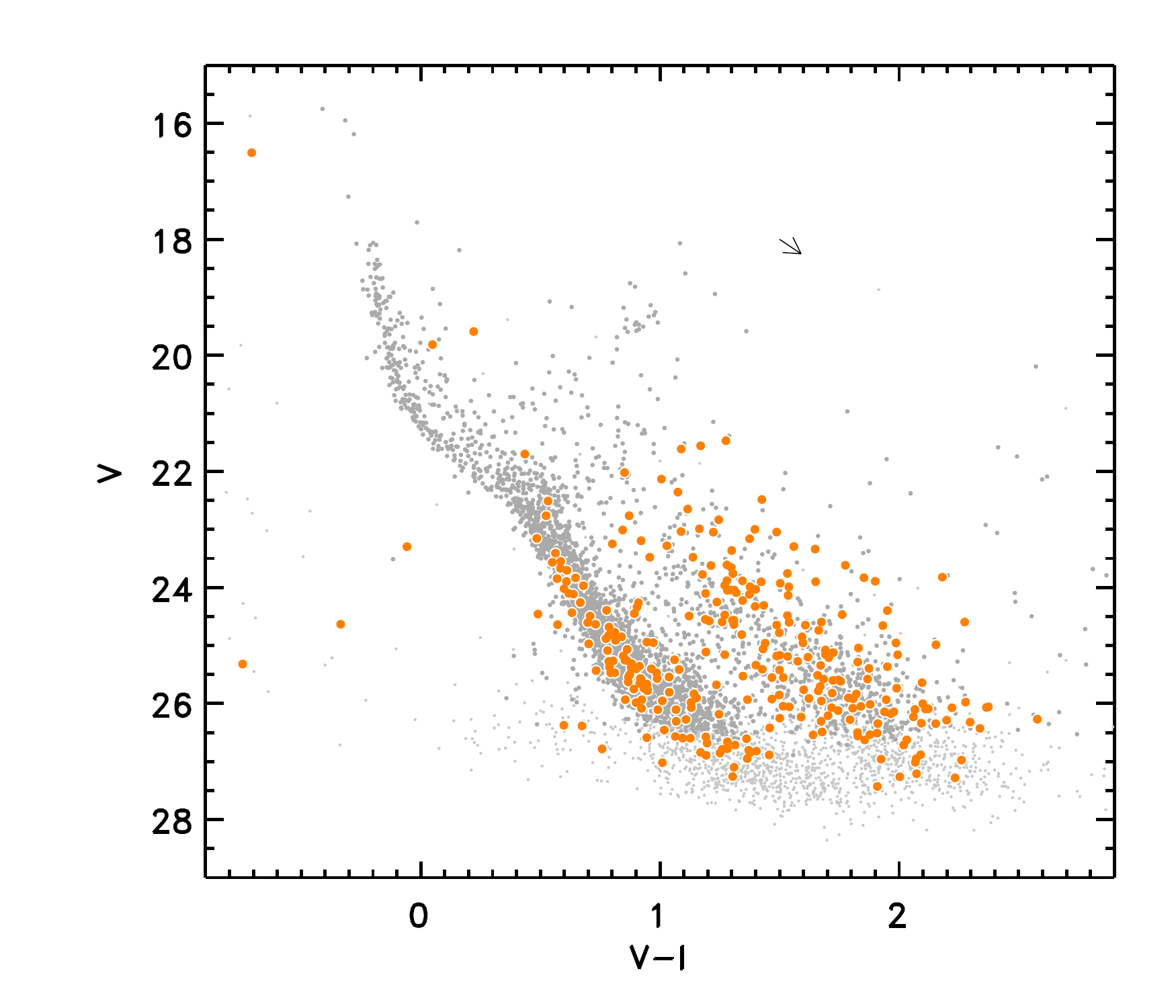}}
\caption{CMD of all stars after correction for reddening. All stars,
except for those indicated by the smallest grey dots, have a combined 
photometric uncertainty in $V$ and $I$ of $0.1$\,mag or less. Objects 
indicated as thick dark dots (orange in the online version) correspond 
to stars with H$\alpha$ excess emission (see Section\,3). } 
\label{fig2}
\end{figure}

The outcome of our photometry is the CMD of the entire field, as shown
in Figure\,\ref{fig2}. Having defined the combined
photometric uncertainty in the $V$ and $I$ bands as:

\begin{equation}
\delta_2=\sqrt{\frac{\delta^2_V+\delta^2_I}{2}},
\end{equation}

\noindent
objects with $\delta_2 < 0.1$\,mag are indicated as larger grey dots,
whereas smaller grey dots are used for the remaining stars (objects
marked as thick dark dots, orange in the online version, correspond to
stars with H$\alpha$ excess emission and are discussed in Section\,3).
The magnitudes shown in the CMD (and in all figures thereafter) have
already been corrected for the effects of interstellar extinction. 
Adopting the canonical value of $E(B-V)=0.08$ (see Carlson et al. 2007
and Schmalzl et al. 2008, and references therein, for an extensive
discussion of the extinction in this region), and assuming that the
reddening law derived by Scuderi et al. (1996) for the field of
SN\,1987A also applies to this region of the SMC, we obtain $A_V=0.25$,
$A_I=0.16$ and $A_{H\alpha}=0.21$ for the specific ACS bands used here.
The corresponding reddening vector is shown by the small arrow in
Figure\,\ref{fig2}.

\section{Searching for pre-main sequence stars}

The CMD of Figure\,\ref{fig2} reveals that in the field covered by these
observations there are stars of different ages: there are young massive
stars in the upper MS (with $V-I < 0$ and $V \la 21$), there are objects
brighter and redder than the MS, in the region usually occupied by PMS
stars, and there are stars in the lower MS and along what appears as the
red-giants branch with a prominent red clump at $V \simeq 19$, which are
usually populated by older objects. This variety of stellar populations
has already been considered and discussed in a number of recent papers
(see Carlson et al. 2007; Schmalzl et al. 2008; Cignoni et al. 2009;
Gouliermis et al. 2012). Note that, owing to the high Galactic
latitude of the SMC, the contamination due to foreground Milky Way
sources is insignificant. According to the Besan\c{c}on Galactic stellar
population synthesis models (Robin et al. 2003), only four foreground
stars with magnitudes in the range $7 < M_V < 20$ should be present 
towards NGC\,602 in a field of the size covered by our observations. 

As mentioned in the Introduction, all these works identify as PMS stars
all objects located appreciably to the right of the MS and above it. In
this paper, however, for the first time we look for the signs of the
active mass accretion process that is expected to accompany the PMS
phase and that is responsible for the strong excess emission normally
observed in objects of this type (e.g. Calvet et al. 2000). In
particular, following the method developed in Papers\,I and II, we have
looked for the presence of an excess in the H$\alpha$ emission line by
using a combination of broad-band ($V, I$) and narrow-band ($H\alpha$)
photometry. This way of identifying PMS stars is more reliable than the
simple classification based on the position of the objects in the CMD or
Hertzsprung--Russell (H--R) diagram and provides for a secure detection
of relatively old PMS stars, already close to the MS.

\begin{figure}[t]
\centering
\resizebox{\hsize}{!}{\includegraphics[width=16cm]{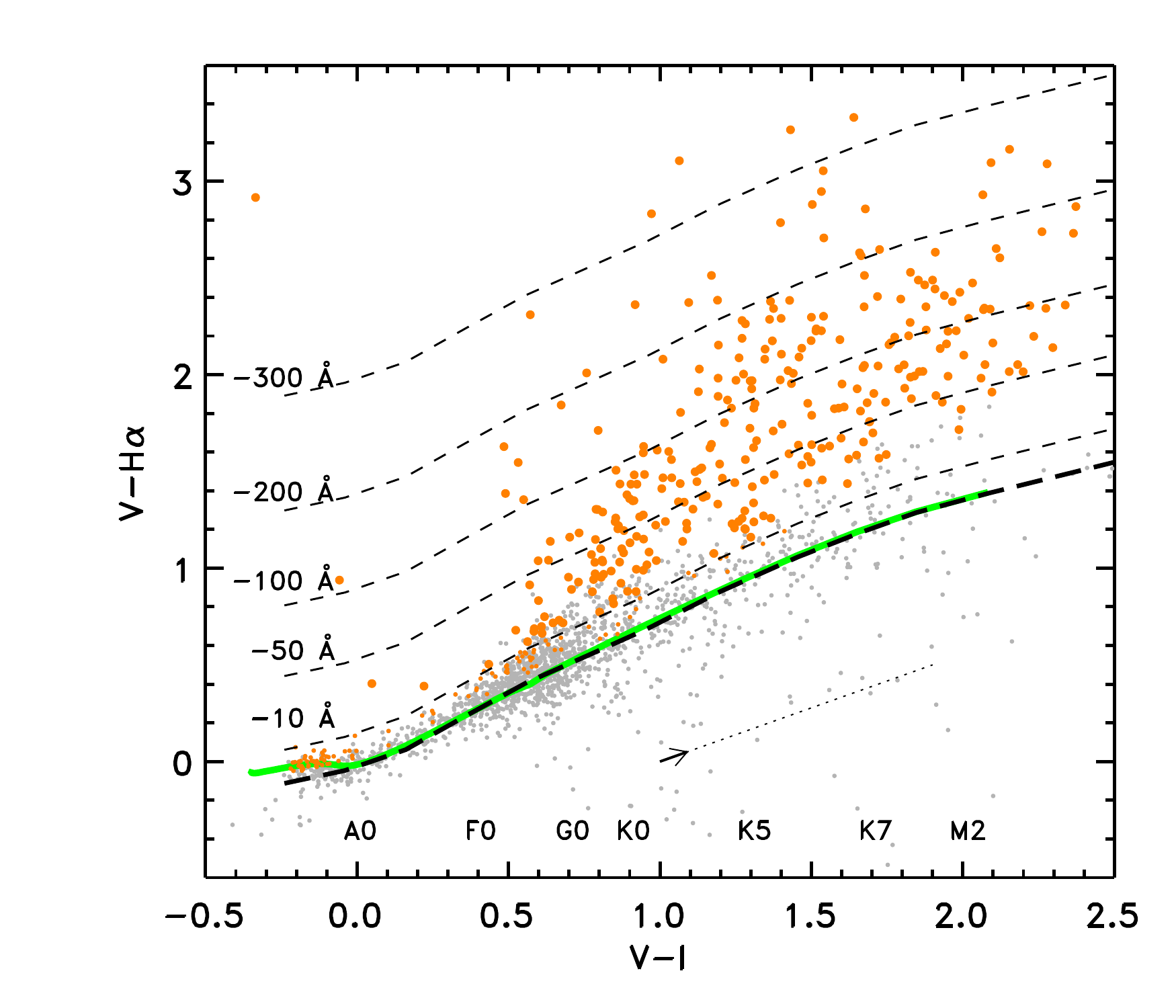}}
\caption{Two--colour diagram of the stars with $\delta_3 \le 0.08$\,mag
(light grey dots), corrected for reddening. The thick long-dashed line
represents the running median $V-H\alpha$ colour, whereas the thick
solid line (green in the online version) shows the model  atmospheres of
Bessell et al. (1998). All objects indicated as large dark dots (orange
in the online version) have a $V-H\alpha$ excess larger than $4\,\sigma$
and an H$\alpha$ excess equivalent width of more than 10\,\AA. The
10, 50, 100, 200, and 400\,\AA\ limits are shown by the short-dashed
lines and labelled accordingly. Objects indicated as
small dark dots (orange in the online version) also have a $V-H\alpha$
excess larger than $4\,\sigma$ but less than 10\,\AA\ for the H$\alpha$
equivalent width. The short arrow shows reddening vector
for $A_V=0.25$, whereas the dotted line extends it by a factor of ten 
purely to show that its direction is almost parallel to that of the
reference template. Spectral types for objects in the range 
$0 < V-I < 2$ are also indicated.}
\label{fig3}
\end{figure}

As explained in Paper\,I, the majority of stars in a typical stellar
field  have no excess H$\alpha$ emission. Therefore, the {\em median}
value of the $V-H\alpha$ colour index at a given effective temperature
$T_{\rm eff}$ defines a spectral reference template for all stars with
that $T_{\rm eff}$ and can be used to identify objects with H$\alpha$
excess. In practice, the $V-H\alpha$ {\em vs.} $V-I$ diagram shown in
Figure\,\ref{fig3} can be used for this purpose. Stars shown as light
grey dots are objects with a combined photometric uncertainty $\delta_3
\le 0.08$\,mag, where

\begin{equation}
\delta_3=\sqrt{\frac{\delta^2_V+\delta^2_{H\alpha}+\delta^2_I}{3}}
\end{equation}

\noindent  
and $\delta_V$, $\delta_{H\alpha}$ and $\delta_I$ are the photometric
uncertainties in the corresponding bands. A total of $1\,906$ objects
satisfy this condition, with the most stringent constraint being set by
the uncertainty in $H\alpha$ (the typical uncertainty in the other two
bands are less than $0.02$\,mag). The log-dashed line represents the
median $V-H\alpha$ colour obtained as the running median with a box-car
size of 100 points (the box-car size is reduced to 10 points for $V-I >
1.5$, where there are fewer stars). It is reassuring to see that the
reference template empirically defined by the dashed line is in
remarkably good agreement with the model atmospheres of Bessell et al. 
(1998), shown by the thick solid line for this specific ACS filter set.

In order to find stars with H$\alpha$ excess emission, it is sufficient
to look for objects with a $V-H\alpha$ index exceeding that of the
reference template at the same $V-I$ in a significant way. As in
previous works (e.g. Papers\,I and II; Beccari et al. 2010; De Marchi et
al. 2011c; Spezzi et al. 2012, hereafter Paper\,III), we set this limit
at the $4\,\sigma$ level and looked for stars that exceed the reference
template by at least four times their $V-H\alpha$ photometric
uncertainty. Indicating with $(V-H\alpha)^{\rm obs}$ the observed value
and with  $(V-H\alpha)^{\rm ref}$ the value on the reference template at
the same $V-I$ colour, we define their difference as 

\begin{equation}
\Delta H\alpha = (V-H\alpha)^{\rm obs} - (V-H\alpha)^{\rm ref}
\end{equation}

\noindent
Then the condition to be satisfied for considering the star a bona-fide
PMS object is that  $\Delta H\alpha > 4\, \sigma_{VH\alpha}$, where
$\sigma_{VH\alpha}$ is the photometric uncertainty on the
$(V-H\alpha)^{\rm obs}$ colour. A total of 400 objects satisfy this
condition and they are shown in Figure\,\ref{fig3} as large dots (orange
in the online version). 

As mentioned above, all magnitudes in Figure\,\ref{fig3} are corrected
for extinction using the average value $A_V=0.25$ for all stars and the
reddening law of Scuderi et al. (1996). The reddening vector for
$A_V=0.25$ is shown by the small arrow, like in Figure\,\ref{fig2}, but
for better illustrating its direction we have  extended it with the
dotted line by a factor of ten, in order to show that the vector is
almost parallel to the reference template. This means that even
relatively large fluctuations in the reddening correction of individual
stars, due for instance to patchy absorption, would not significantly
affect the identification of stars with H$\alpha$ excess.

All these stars have a robust H$\alpha$ excess emission and as such they
are candidate PMS objects. However, to further restrict the list to the
most certain PMS stars, following White \& Basri (2003), Paper\,I and
Barentsen et al. (2011), we conservatively limit our selection to stars
which also have an equivalent width of the H$\alpha$ emission line of
more than 10\,\AA, or $W_{\rm eq}(H\alpha) < -10$\,\AA\ if, as
customary, negative values of the equivalent width are used for emission
lines. According to White \& Basri (2003), $W_{\rm eq}(H\alpha) <
-10$\,\AA\ is a clear sign of accretion in genuine T Tauri stars of
spectral type earlier than M\,2.5 (roughly $V-I = 2.3$ for the specific
ACS filters), since it cannot be caused by chromospheric activity. 

As regards the equivalent widths, we derive them directly from the
colour difference between the $V-H\alpha$ index of the star and that of
the reference template in Figure\,\ref{fig3} and from the properties of
the $H\alpha$ filter, as explained in detail in Paper\,I. More
precisely, if the stars defining the reference template had no H$\alpha$
absorption features, their $V-H\alpha$ colour would correspond to that
of the pure continuum and could be directly used to derive the
equivalent width of the H$\alpha$ emission line, $W_{\rm } (H\alpha)$.
As explained in Paper\,I, although certainly valid for stars with large
H$\alpha$ excess, this assumption is not true in general. Instead, given
the excellent agreement between the reference template and the model
atmospheres of Bessell et al. (1998), we can use the a modified version
of the latter, with the H$\alpha$ emission line suppressed, to derive
the magnitude of the sole continuum in the specific $H\alpha$  band of
the ACS, as a function of spectral type or $V-I$ colour. Since the
H$\alpha$ line profile is very narrow when compared to  the width of
the filter, the equivalent width is simply given by the relationship:

\begin{equation}
W_{\rm eq} (H\alpha) =  \mathrm{RW} \times [1-10^{-0.4 \times
(H\alpha-H\alpha^c)}]
\label{eq4}
\end{equation}

\noindent
where $\mathrm{RW}$ is the rectangular width of the filter (similar in
definition to the equivalent width of a line), which depends on the
characteristics of the filter, $H\alpha$ is the measured magnitude and 
$H\alpha^c$ the magnitude of the sole continuum. The value of
$\mathrm{RW}$ for the ACS F658N filter used here is $74.98$\,\AA\ and 
values for the $H\alpha$ bands of current and past HST instruments are
given in Paper\,I. 

A small correction is then applied to the derived $W_{\rm eq} (H\alpha)$
values to account for the fact that for normal stars (those in grey in
Figure\,\ref{fig3}) the H$\alpha$ line is normally in absorption (see
Paper\,I). The stars with $W_{\rm eq}(H\alpha) < -10$\,\AA\ are shown as
large orange dots in Figure\,\ref{fig3}, and are located above the short
dashed line corresponding to $W_{\rm eq}(H\alpha) =-10$\,\AA. In 
total we find 296 objects redder than $V-I=0$ that satisfy this
condition, with a median equivalent width $W_{\rm eq}(H\alpha) \simeq
-60$\,\AA\ and a median H$\alpha$ luminosity $L(H\alpha) \simeq 2.4
\times 10^{-3}$\,L$_\odot$. Hereafter, we take these objects as
bona-fide PMS stars. The median $W_{\rm eq}(H\alpha)$ and
$L(H\alpha)$ values might appear larger than those typical of Galactic
T-Tauri stars, but as explained above our sample only includes objects
with $W_{\rm eq}(H\alpha) < 10$\,\AA\ and therefore stars with low
$L(H\alpha)$ are automatically excluded.

These objects are shown in the CMD of Figure\,\ref{fig2} as large
dots (orange in the online version). Interestingly, not all of them are
in the region occupied by young PMS stars, i.e. brighter and redder than
the MS, as some are very close to or at the MS itself. This is a
consistent feature of all the star-forming regions previously observed
by us in the Magellanic Clouds and in the Milky Way (e.g. Panagia et al.
2000; Papers\,I and II; Beccari et al. 2010; De Marchi et al. 2011c;
Paper\,III). Several types of H$\alpha$ emitting sources
could occupy the MS region the CMD but, as we will now show, the
most likely explanation is that these are older PMS stars, approaching
the MS and  still actively accreting. 

Examples of objects with H$\alpha$ excess in this region of the CMD 
include cataclysmic variables (CVs) and stars with active chromospheres,
such as RS CVn objects or dMe stars. Chromospheric activity, however,
can be immediately ruled out, because it causes very weak H$\alpha$
emission for stars earlier than spectral type M2, with equivalent widths
typically of order $-1$\,\AA\ (e.g. Bopp \& Talcott 1978; Bopp \&
Schmitz 1978). Precisely, it is to keep our sample free from objects of
this type that, following White \& Basri (2003), we only consider stars
with $W_{\rm eq}(H\alpha) < -10$\,\AA. 

On the other hand, CVs have typical $W_{\rm eq}(H\alpha)$ in the same
range as the one we measure (e.g. Warner 1995 and references therein),
but these objects are very infrequent. Pretorius \& Knigge (2008)
searched  for objects with H$\alpha$ excess in the $R-H\alpha$ vs. $R-I$
diagram  (similar to our Figure\,\ref{fig3}), using observations from
the AAO/UKST  SuperCOSMOS H$\alpha$ Survey (Parker et al. 2005). They
analysed the $R$ and $H\alpha$ photometry down to $R=17$ in 175
regions,  each 4\,deg on a side, for a total covered area of $\sim
2\,800$\,deg$^2$ at low Galactic latitudes ($\vert b \vert < 2$\,deg).
Each region contains of order 170\,000 stars, for a total of about 30
million objects, and only 14 CVs were found amongst them, with an
implied frequency of order $5 \times 10^{-7}$. Considering that the
typical distance to these CVs is  $\sim 500$\,pc (Pretorius \& Knigge
2008), the $R=17$ magnitude limit for  these stars  translates to about
$R=27.5$ at the distance of NGC\,602 and as such it is very similar to
the detection limit of our photometry. Therefore, with a total of 5\,500
stars in our field we should expect essentially no ($0.003$) CVs. 

Considering that the metallicity of the stars in the SMC is about a
factor of five lower than that of the objects in the Galactic plane, we 
have also looked at the fraction of CVs in Galactic globular clusters
(GC),  which have lower metallicity. The cluster 47\,Tuc, with
$[Fe/H]=-0.76$ (Harris 1996), offers conditions very similar to those of
the SMC field. This cluster contains the largest known number of CVs in
a GC (Downes et al. 2005), most of which are located near the cluster
centre (Heinke et al. 2005). Inside a field of $3\farcm4 \times
3\farcm4$ around the cluster centre, observed with the HST as part of
the ACS Globular Clusters Survey (Sarajedini et al. 2007), there are in
total $\sim 100\,000$ stars down to $V \simeq 24$. Of these stars, 30
are classified as CVs from their X-ray properties (Heinke et al. 2005) and
only six of them have excess emission in H$\alpha$ with  $W_{\rm
eq}(H\alpha) < -20$\,\AA\ (Beccari et al. 2013), with an implied
frequency of $6 \times 10^{-5}$. Therefore, with a total of 5\,500
stars, the expected number of CVs in our field ($0.3$) remains
insignificant. 

In summary, it is possible (and perhaps even likely) that some of the
stars with H$\alpha$ excess bluer than the MS in Figure\,\ref{fig2} are
objects of this type. Precisely for that reason we do not consider any
objects with H$\alpha$ excess bluer than $V-I=0$ in this work. However,
it is clear that CVs cannot be responsible for the large population of
objects with H$\alpha$ excess near the MS. As concluded in the previous
works in this series, the most likely nature of these objects is that
they are recently formed stars still actively accreting while they
approach the MS. In the next section we will derive the physical
properties of these and of the younger PMS stars through comparison with
PMS evolutionary models.

\section{Physical parameters of the PMS stars}

\begin{figure}[t]
\centering
\resizebox{\hsize}{!}{\includegraphics[width=16cm]{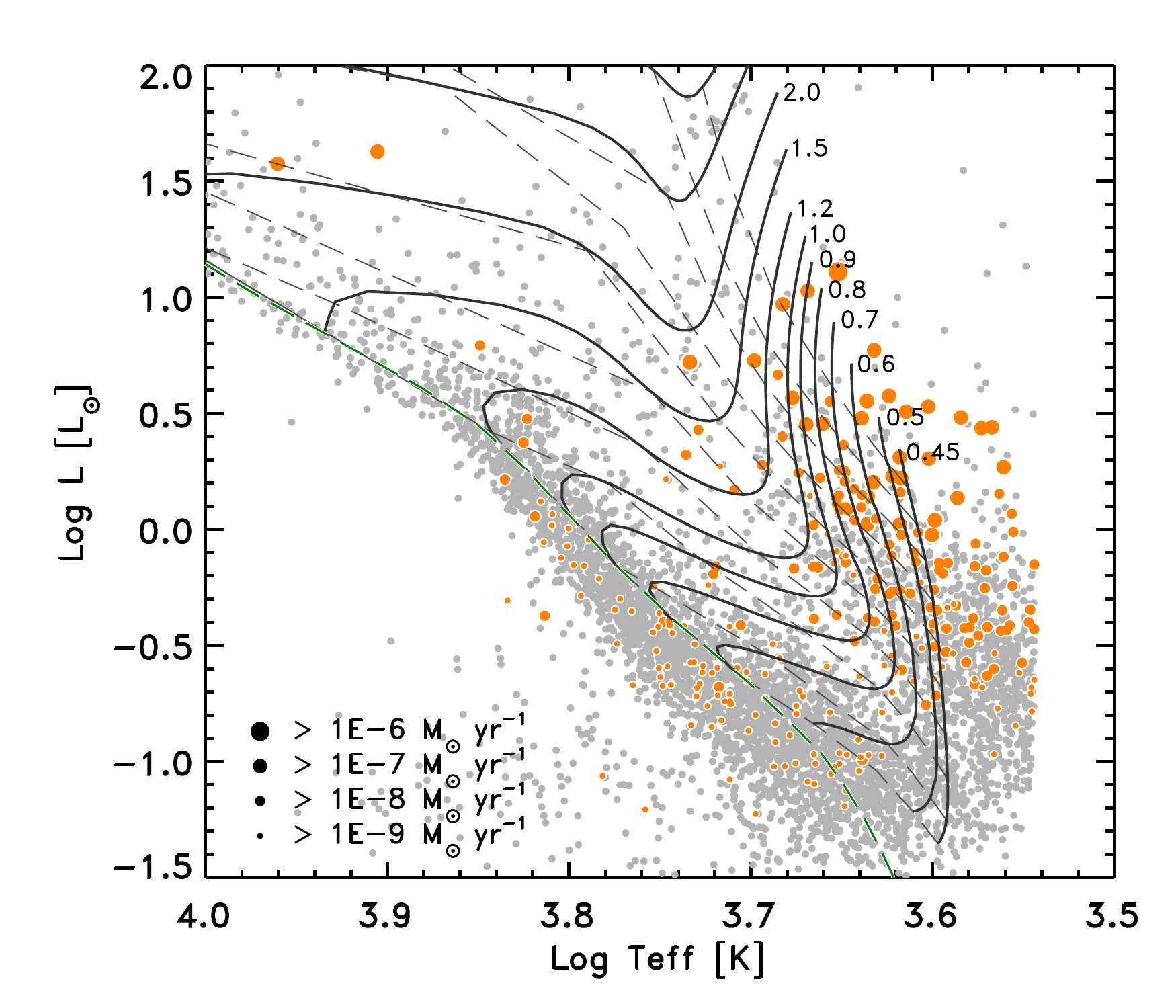}}
\caption{All objects  with combined photometric uncertainty smaller than
$0.1$\,mag in $V$ and $I$ are shown in grey in the H--R diagram.
Bona-fide  PMS stars are shown as dark dots (orange in the online
version), with a symbol size proportional to their mass accretion rate,
as per the legend. Thick  solid lines show the evolutionary tracks for
metallicity $Z=0.004$ and masses as indicated. The corresponding 
isochrones are shown as thin dashed lines, for ages of $0.125$, $0.25$,
$0.5$, 1, 2, 4, 8, 16, 32, and 64\,Myr from right to left. The constant
logarithmic age step has been selected in such a way that the typical
distance between isochrones is larger than the photometric
uncertainties. The leftmost dashed line (green in the online
version) is the ZAMS from the models of Marigo et al. (2008) for
the assumed $Z=0.004$ metallicity.} 
\label{fig4}
\end{figure}

The luminosities ($L$) and effective temperatures ($T_{\rm eff}$) of all
stars with photometric uncertainty in $V$ and $I$ smaller than
$0.1$\,mag are shown in the H--R diagram (Figure\,\ref{fig4}) as light
grey dots, whereas PMS stars are shown as dark dots (orange in the
online version; note that the size of the dots is proportional to the
mass accretion rate, as explained later).  Effective temperatures are
determined by fitting the dereddened $V-I$ colours of the stars with the
ones computed for the same HST bands using stellar  atmosphere models.
We have adopted the model atmospheres of Bessell et al. (1998) for the
metallicity $[M/H]=-0.5$ appropriate for the SMC, as they provide a
well-tested, homogeneous set covering the temperature interval between
3\,500 and 50\,000 K in the wavelength range between 90\,\AA\ and
160\,$\mu$m. The dereddened magnitudes are then compared with the
absolute magnitudes of the models of Bessell et al. (1998) computed for
the same bands. These models were renormalised by Romaniello et al.
(2002; see their Table 6) in order to provide the absolute magnitudes of
stars of a given temperature and surface gravity, for a radius of
1\,R$_\odot$ and a distance of 10\,pc. Knowing the distance to each
source, we can derive its radius $R_*$ since the magnitude difference
$\Delta M_V$ scales as  $\Delta M_V = 5 \times \log (R_*/R_\odot)$. 
Finally, the bolometric luminosity is derived knowing the distance to
the objects, from the stellar radius and effective temperature, compared
to those of the sun (see Romaniello et al. 2002 for further details). As
for the distance to the SMC, we have assumed a distance modulus of
$18.92 \pm 0.03$ corresponding to $\sim 61$\,kpc (Hilditch et al. 2005;
Keller \& Wood 2006). Note that the sharp cut-off at $\log T_{\rm
eff}=3.55$ in Figure\,\ref{fig4} is caused by the limited extent of the
model atmospheres and, therefore, all objects redder than $V-I \simeq
2.1$ are ignored.

We also show in Figure\,\ref{fig4} the evolutionary tracks
(thick solid lines) and isochrones (dashed lines) from the PMS
evolutionary models of the Pisa group (Degl'Innocenti et al. 2008;
Tognelli et al. 2012). The models have been calculated for $Z=0.004$,
$Y=0.24$ and the mixing-length parameter $\alpha = 1.9$ for masses in
the range $0.45 - 5.5$\,\Msolar. As shown by Cignoni et al. (2009), with
the assumed distance modulus and reddening $E(B-V)\simeq 0.08$ and this
choice of the metallicity, the Padua isochrones (Fagotto et al. 1994;
Marigo et al. 2008) provide an excellent fit to the upper MS. This
metallicity also nicely agrees with the currently accepted values for
the SMC, which range from $\sim 1/5$ to $\sim 1/8$\,Z$_\odot$ (see
Russell \& Dopita 1992; Rolleston et al. 1999; Lee et al. 2005;
Perez--Montero \& Diaz 2005). 

It should be noted, however, that this metallicity is a factor of two
higher than that implied by the observations of NGC\,346, in the bar of
the SMC, where a metallicity value of $Z=0.002$ is preferable (see
Paper\,II). This difference is also consistent with the available
measurements of the ratio between extinction and column density of
neutral hydrogen $A_V/N(H\,I)$. From Gordon et al. (2003), expressing
$N(H\,I)$ in units of $10^{22}$\,cm$^{-2}$, we find the ratio to be $1.35
\pm 0.22$ in the SMC wing and a value about a factor of two lower in the
SMC bar ($0.76 \pm 0.06$). Assuming that the ratio scales linearly with
metallicity (e.g. Scuderi et al. 1996), this result suggests indeed a
factor of $\sim 2$ higher metal content in NGC\,602 than in NGC\,346.

In Figure\,\ref{fig4}, the masses corresponding to the individual tracks
are indicated on the right-hand side, whereas the ages of the isochrones
increase by a factor of two at each step, starting from $0.125$\,Myr at
the right end to 64\,Myr at the left end. We also show for comparison
the zero age MS of Marigo et al. (2008) for the same metallicity
(leftmost dashed line, in green in the online version). 

\begin{figure}[t]
\centering
\resizebox{0.8\hsize}{!}{\includegraphics{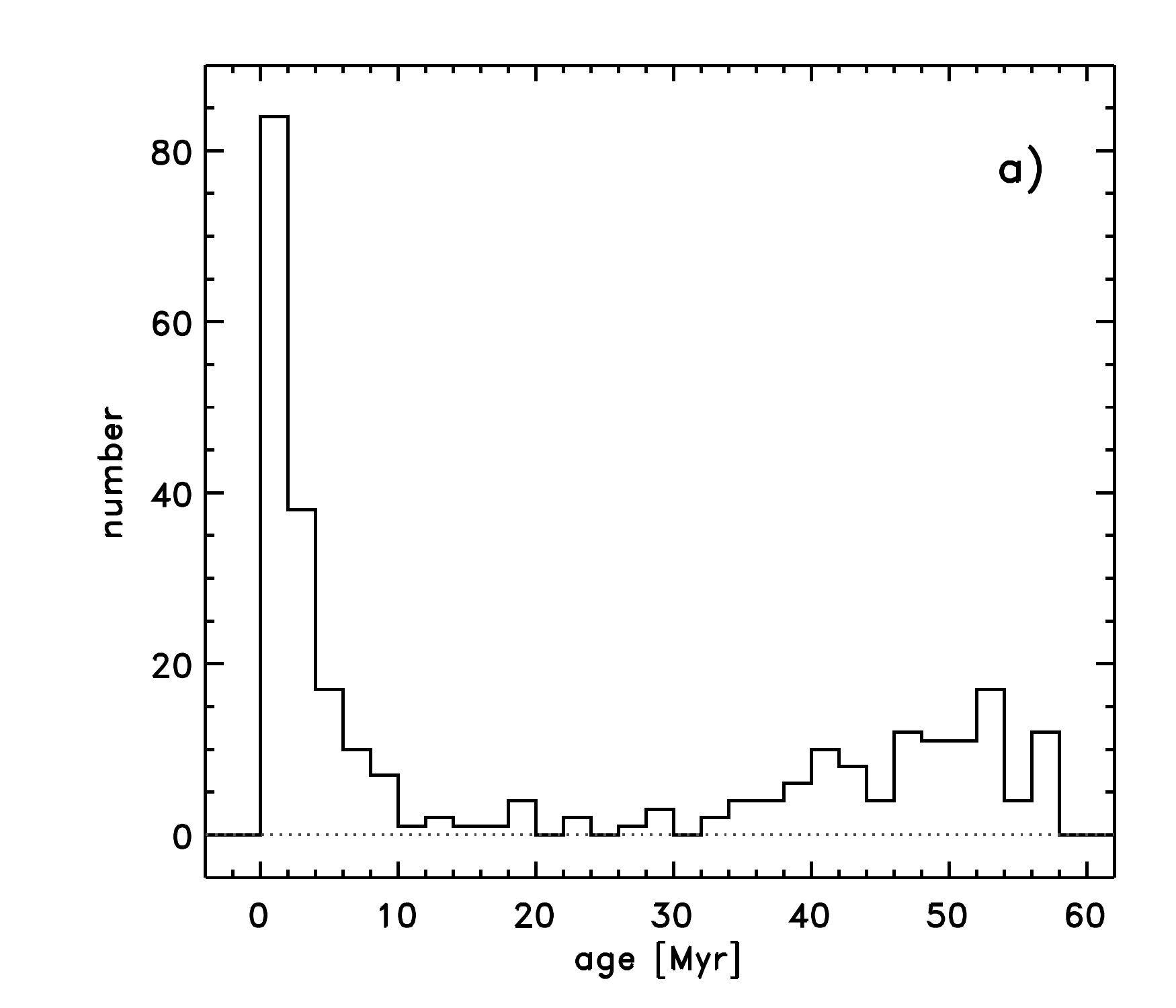}}
\resizebox{0.8\hsize}{!}{\includegraphics{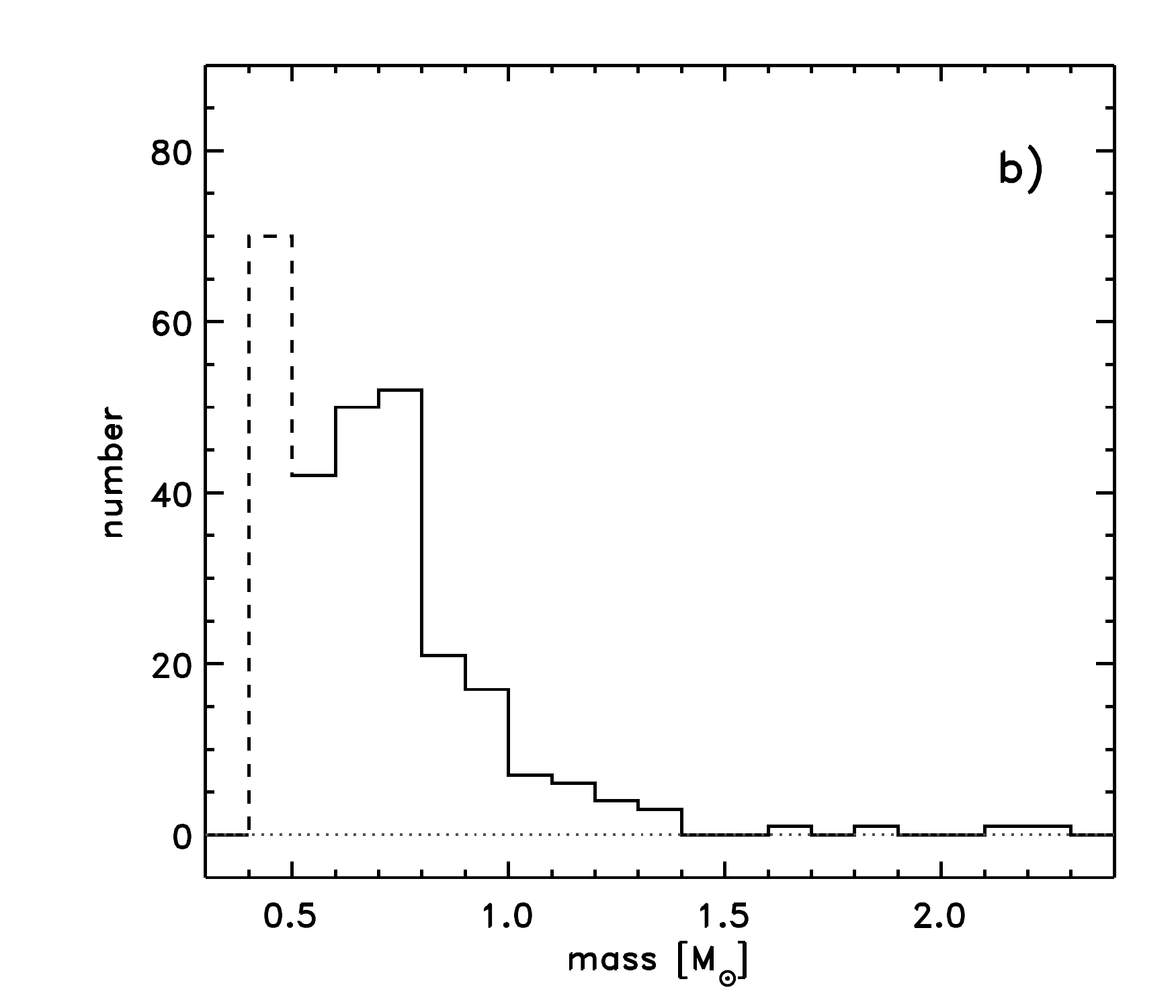}}
\caption{Age and mass distribution of PMS stars. The first mass bin is
shown with a dashed line because it could also include  stars of lower
mass.} 
\label{fig5}
\end{figure}

The comparison with the Pisa models immediately suggests ages older than
$\sim 32$\,Myr for most of the PMS stars close to the MS, ages younger
than $\sim 4$\,Myr for the rest and few objects in between. Concerning
the masses, the majority of PMS stars appear to be less massive than
$\sim 0.8$\,\Msolar. Using a finer grid of models than the one shown in
Figure\,\ref{fig4}, we derived a more accurate measure of the mass and
age for 276 PMS stars, excluding the 20 coolest objects that would have
required a rather uncertain extrapolation beyond the limits of the
theoretical models (see Romaniello 1998 for details on the method used,
with an approach similar to the one recently presented by Da Rio et al.
2012). While an accurate determination of the absolute age and mass of
individual objects will require spectroscopy only possible with future
instrumentation (e.g. the James Webb Space Telescope; Gardner et al.
2006), we can nonetheless derive important information on the properties
of star formation in this field by comparing relative ages and masses of
groups of stars. Histograms with the age and mass distribution for the
276 stars are shown in Figures\,\ref{fig5}a and \ref{fig5}b,
respectively. Besides the remarkable paucity of stars in the $\sim
10-30$\,Myr range, already seen in the H--R diagram, Figure\,\ref{fig5}a
shows that about {\small 1/3} of our PMS objects have isochronal ages
older than $\sim 30$\,Myr. The double peak seen in the figure implies
that in the past this region underwent at least one and possibly several
bursts of star formation, likely as intense as the one currently
revealed by the 0--2\,Myr old objects. We will come back to this point
in Section\,5.

The $\Delta H\alpha $ parameter defined in the previous section also
allows us to easily derive the H$\alpha$ emission line luminosity
$L(H\alpha)$ for these stars. As explained in Paper\,I, $L(H\alpha)$ is
obtained from $\Delta H\alpha$, from the photometric zero point and
absolute sensitivity of the instrumental set-up and from the distance to
the sources. As for the latter, we have assumed a distance modulus
$(m-M)_0=18.92$ as mentioned before, whereas the photometric properties
of the instrument are as listed in the ACS Instrument Handbook (Maybhate
et al. 2010). The median $L(H\alpha)$ in our sample is $\sim 2 \times
10^{-3}$\,L$_\odot$, with the 17 and 83 percentile levels in the
distribution at $0.001$\,L$_\odot$ and $0.01$\,L$_\odot$, respectively.
We stress here that, although our sample only includes objects with
$W_{\rm eq}(H\alpha) < -10$\,\AA, the $L(H\alpha)$ values that we find
are in excellent agreement with those of Galactic stars of similar 
mass. Dahm (2008) analysed the H$\alpha$ luminosity of a group of T
Tauri stars in Taurus--Auriga, taken from the original sample of Valenti
et al. (1993), with masses similar to those of the objects in our
sample. He obtains $L(H\alpha)$ luminosities ranging from
$0.001$\,L$_\odot$ (17 percentile) to $0.014$\,L$_\odot$ (83
percentile), 
in excellent agreement with our measurements.    

The uncertainty on $L(H\alpha)$ is typically $\sim 12$\,\% and is
completely dominated by the statistical uncertainty on the $H\alpha$
photometry (10\,\%), with the distance accounting for 5\,\% (Hilditch et
al. 2005; Keller \& Wood 2006) and the absolute sensitivity of the
instrumental setup for another 3\,\%. These values are corrected for the
small contribution from the [NII] emission features at $6\,548$\,\AA\
and $6\,584$\,\AA\ that are included in the rather wide F658N passband
of the ACS. The correction is very small, only $2.1$\,\%, but since it
is a systematic effect it is a good practice to apply it.

Having measured $L(H\alpha)$ we can derive another important physical 
parameter for the PMS stars, namely the accretion luminosity $L_{\rm
acc}$. As discussed in Paper\,I, Dahm's (2008) logarithmic best fit to
the $\log L_{\rm acc}$ {\em vs.} $\log L(H\alpha)$ plane would provide a
slope of $1.18 \pm 0.26$ for such a relationship. On the other hand,
theoretical models (e.g. Muzerolle et al. 1998) would predict
logarithmic slopes of about unity for low accretion rates, i.e. faint
$L(H\alpha)$, and shallower slopes for higher luminosities. In the
absence of compelling evidence in favour or against a steep slope, we
adopt a logarithmic slope of unity for the empirical $\log L_{\rm acc}$ 
{\em vs.} $\log L(H\alpha)$ relationship, which is inside the 
uncertainty band on Dahm's (2008) value. In their recent study of
accretion rates for T Tauri stars using nearly simultaneous ultraviolet
and optical observations, Ingleby et al. (2013) find consistently a
slope of unity. Not only for the luminosity in H$\alpha$ ($1.0 \pm
0.2$),  but also for other important accretion indicators, namely the
luminosity in $H\beta$ ($1.0 \pm 0.1$), Ca II K ($1.0 \pm 0.2$), C II]
($1.1 \pm 0.2$) and Mg II ($1.1 \pm 0.2$). A choice of unity has a
physical meaning and implies that a constant fraction of the
gravitational energy released in the accretion process goes into
H$\alpha$ emission. In Paper\,I and II we determined the proportionality
constant from the data summarised by Dahm (2008),  combined with the
dispersion measured for the stars in NGC\,346, finding

\begin{equation}
\log L_{\rm acc} = (1.72 \pm 0.25) + \log L(H\alpha).
\end{equation}


We will make use of $L_{\rm acc}$ obtained in this way in Section\,6 for
deriving the mass accretion rate $\dot M_{\rm acc}$. The median
value of $L_{\rm acc}$ that we find for the PMS stars in  our sample is
$\sim 0.1$\,L$_\odot$. Comparing this value to the typical bolometric
luminosities $L_{\rm bol}$ of these objects, we find the   median
$L_{\rm acc} / L_{\rm bol}$ ratio to be $\sim 0.3$. The 17 and 83
percentiles, which in a Gaussian distribution would correspond to the
$\pm 1\,\sigma$ values, are respectively $0.15$ and $0.63$. The derived
$L_{\rm acc} / L_{\rm bol}$ ratio indicates that in our sample the
accretion luminosity does not dominate the total stellar luminosity, as
is expected of stars that are no longer in the embedded phase (e.g.
Nisini et al. 2005; Antoniucci et al. 2008). Taking again Valenti et
al.'s (1993) sample of PMS stars in Taurus--Auriga, the median $L_{\rm
acc} / L_{\rm bol}$ ratio is $\sim 0.15$, whereas the corresponding
value in our sample is $\sim 0.3$ or a factor of two higher. However, as
explained above, we restrict our analysis to stars with $W_{\rm
eq}(H\alpha) < -10$\,\AA, thereby excluding all objects with small
H$\alpha$ excess and small H$\alpha$ luminosity, and therefore it is to
be expected that we will miss objects with small accretion luminosity.

\section{Star formation across time and space}

As discussed in Section\,3, the age distribution of the PMS stars in our
sample is bimodal (see Figure\,\ref{fig5}a). About {\small 1/2} of the
objects appear younger than $\sim 5$\,Myr and about {\small 1/3} older
than $\sim 30$\,Myr, with few objects (50) at intermediate ages. These
ages are based on the comparison of the observations with theoretical
isochrones and, as such, they are subject to uncertainties. Besides
photometric errors and uncertainties in the  input physics affecting the
models, there are other physical effects that could cause an incorrect
determination of the age (or mass) of individual objects. These include
for instance unresolved binaries, differential reddening, stellar
variability, veiling resulting from accretion, and scattering due to a
disc seen at high inclination. All these effects combine to produce a
broadening in the H--R diagram, which could be misinterpreted as an age
spread (see e.g. Hennekemper et al. 2007 and Da Rio et al. 2010). 


\begin{figure}[t]
\centering
\resizebox{0.9\hsize}{!}{\includegraphics{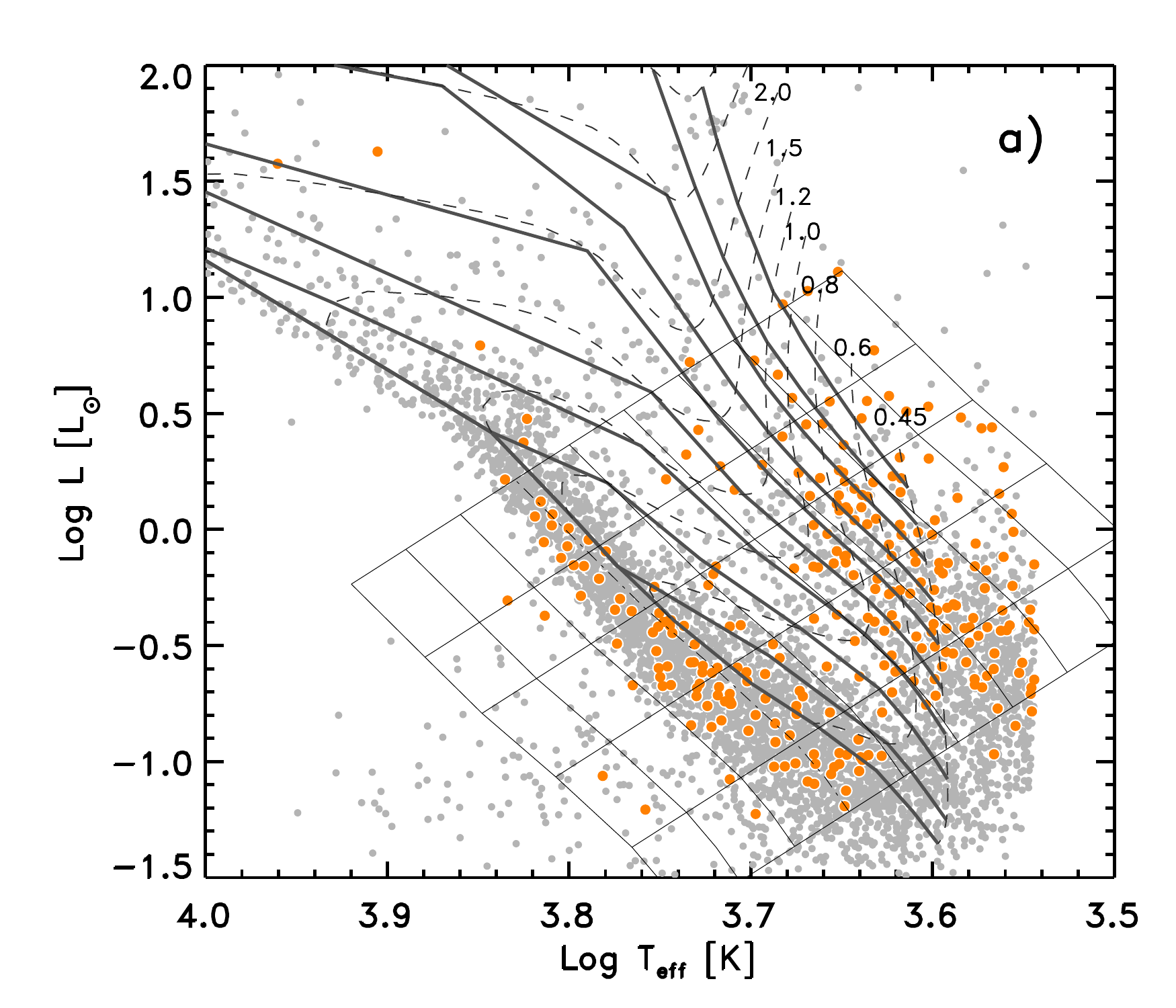}}
\resizebox{0.9\hsize}{!}{\includegraphics{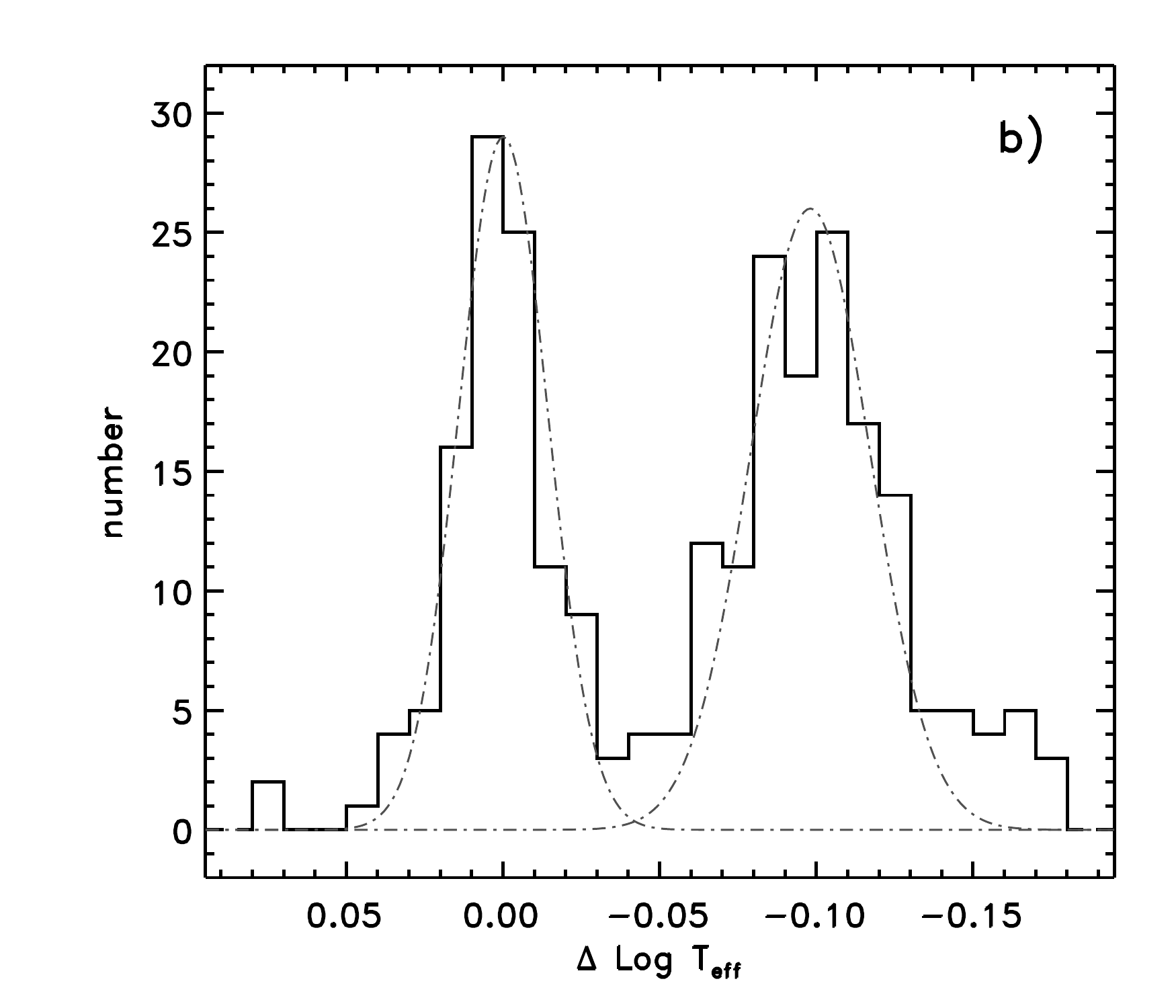}}
\caption{Distribution of PMS stars in the H--R diagram obtained by 
counting the number of objects falling in strips parallel to the ZAMS.
The grid shown in panel a) corresponds to $\Delta \log T_{\rm
eff}=0.03$, for display purposes, but the histograms in panel b) are
obtained using a finer grid ($\Delta \log T_{\rm eff}=0.01$). The value
of $\Delta \log T_{\rm eff}=0$ corresponds to the dot-dashed line in
panel a). The thin dot-dashed lines in panel b) show a Gaussian fit to
the two peaks.} 
\label{fig6}
\end{figure}

In fact, even though all the effects mentioned above could mimic an age
spread, none of them can produce the clearly bimodal distribution that
we observe in the H--R diagram. The remarkable paucity of PMS stars with
age comprised between the 4\,Myr and 16\,Myr isochrones (already evident
in Figure\,\ref{fig4}) is quantified in Figure\,\ref{fig6}, where we
show the distribution of PMS stars as a function of the difference in
their $\log T_{\rm eff}$ values. The distribution is obtained by
counting the number of PMS stars in strips parallel to the zero-age MS
(ZAMS), starting from the dot-dashed line, and it is clearly bimodal,
with two peaks separated by several times their width. A Gaussian fit to
the two peaks (dot--dashed lines in Figure\,\ref{fig6}b) gives
$\sigma_1=0.020$\,dex for older PMS stars and $\sigma_2=0.025$\,dex for
younger PMS stars.  The separation between the two peaks ($0.1$\,dex)
corresponds to respectively 5 and 4 times these widths and confirms that
the two distributions are clearly distinct.

Physically, it is hard to imagine a mechanism that would affect the
temperatures and luminosities of stars in such a way that they are
selectively displaced from the region occupied by young PMS objects in
the H--R diagram and moved towards the MS, while leaving only a handful
of them in the region in between. 
Therefore, while the effects mentioned above can introduce uncertainties
on the relative ages of individual objects (in Paper\,I these are
estimated to be typically no more than a factor of $\sqrt{2}$), we
can safely conclude that the two groups of stars with H$\alpha$ excess
seen in the H--R diagram must belong to different generations, with ages
that differ by much more than a factor of two and likely up to an 
order of magnitude. 

\begin{figure}[t]
\centering
\resizebox{\hsize}{!}{\includegraphics[width=16cm]{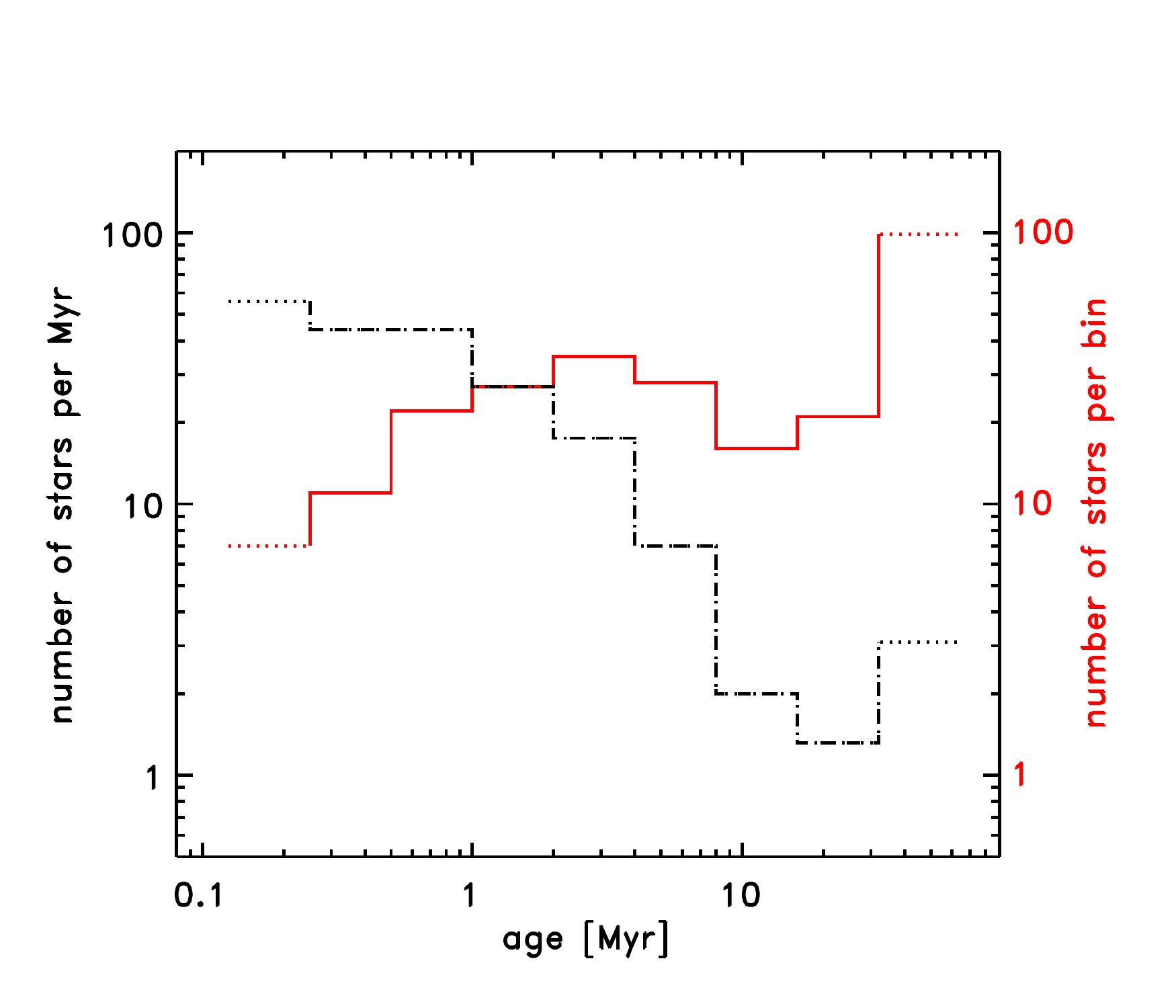}}
\caption{Histograms showing the number of stars per age bin (thick solid
line, red in the online version)  and the apparent star formation
rate, i.e. the number of stars per Myr (dot-dashed line), as a function
of age. Only bona-fide PMS stars with H$\alpha$ excess emission in the
range $0.4-2.5$\,\Msolar\, are considered.} 
\label{fig7}
\end{figure}

\subsection{Multiple stellar generations}

A histogram of the age distribution was already shown in
Figure\,\ref{fig5}a.  Considering the uncertainties inherent in our
age determination, stemming from the comparison of model isochrones with
our photometric data, the histogram in Figure\,\ref{fig7} offers a more
realistic representation of the relative age distribution, with ages
binned using a constant logarithmic step (a factor of 2). The solid
line gives the actual number of stars contained within each age bin,
while the dot-dashed line provides a measurement of the apparent star
formation rate, obtained by dividing the number of objects in each bin
by the width of the bin itself (we use a dotted line at the extremes of
the distribution to indicate larger uncertainties).

Note that the dot-dashed line in Figure\,\ref{fig7} clearly provides a
lower limit to the star formation rate. First of all, crowding effects,
particularly in the proximity of massive stars, make the detection of
faint low-mass PMS stars more difficult. Furthermore, here we only
consider PMS stars in the range $0.4 - 2.5$\,\Msolar\, having H$\alpha$
excess emission at the $4\,\sigma$ level or more at the time of the
observations.  Since the strength and equivalent width of the 
H$\alpha$ emission line are highly variable in time (e.g. Fernandez et
al. 1995; Smith et al. 1999; Alencar et al. 2001), the objects that at
any given time show H$\alpha$ excess emission at these levels are
necessarily only a fraction of the total population of PMS stars. 

Assuming that all objects cooler than the 5\,Myr isochrone in the
H--R diagram of Figure\,\ref{fig6}a are PMS stars, we find that the
fraction of those with H$\alpha$ excess emission is $\sim 15$\,\%.
Therefore, for young stars, the upper limit to the number of PMS objects
is $\sim 6.5$ times larger than what we measure. As regards older PMS
stars, one expects the fraction of objects with H$\alpha$ excess to be
lower than for younger stars. Indeed, as we will show in Section\,5.3,
for two older clusters in the field this fraction is $\sim 8$\,\% along
the MS in the range $24 < V < 27$. Thus, the actual number of PMS stars
for ages older than $\sim 20$\,Myr is at most $\sim 12.5$ times larger
than what we measure, since the stars without excess along the MS may
include much older field stars. 

\begin{figure*}[t]
\centering
\resizebox{\hsize}{!}{\includegraphics[width=16cm]{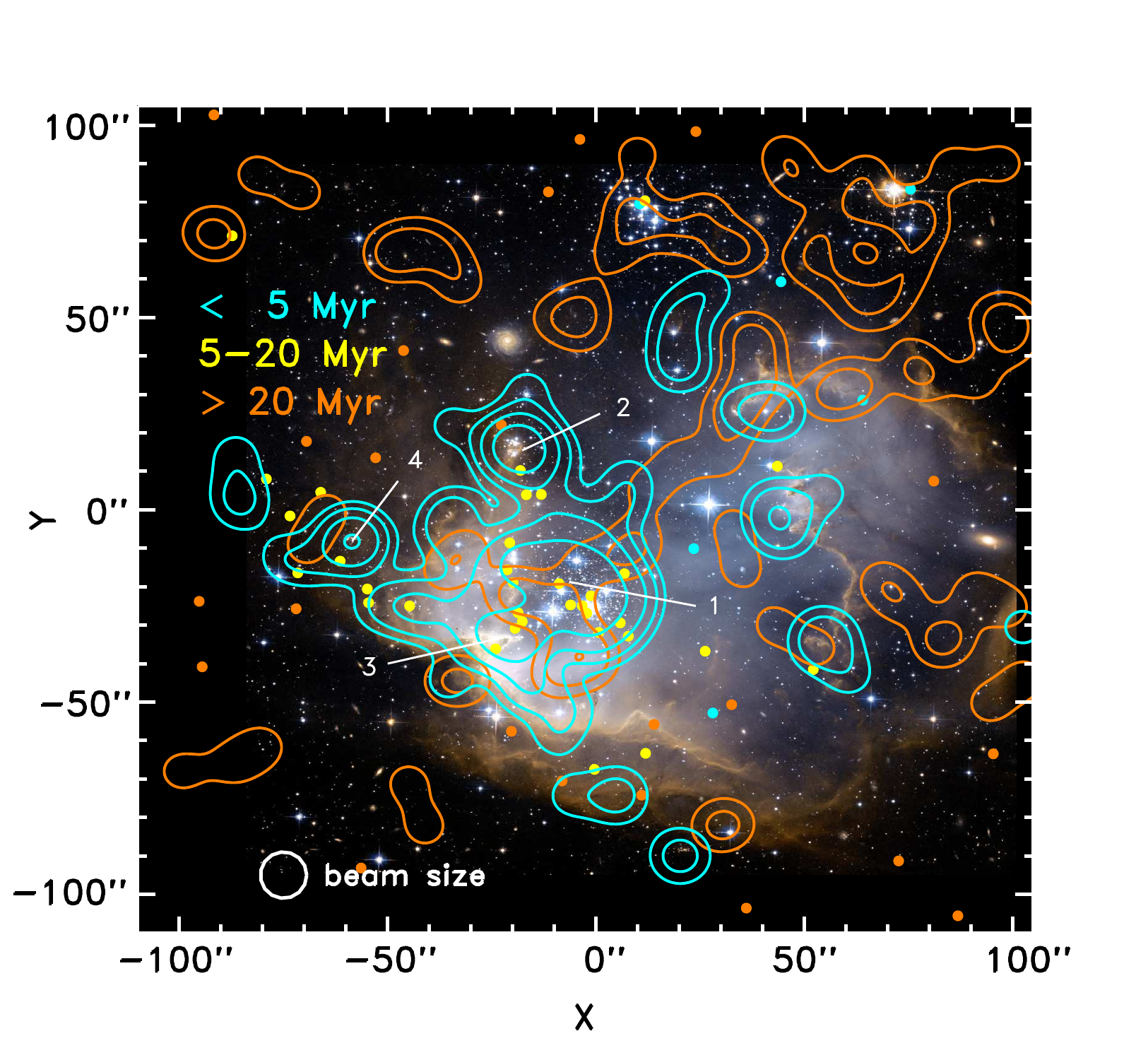}}
\caption{Contour plots showing the position and density distribution of
PMS stars of different ages, as per the legend, overlaid on a
true-colour image of the region. The lowest level corresponds to a
density three times as high as the average density of PMS stars in the
field. The step between contour levels is constant and corresponds to a
factor of 2. Numbers 1 through 4 indicate the positions of the regions
whose properties are shown in Table\,\ref{tab1}.} 
\label{fig8}
\end{figure*}

Near the peak of the distribution in Figure\,\ref{fig7}, at $\sim
0.5$\,Myr, the lower limit to the star formation rate is $\sim 3 \times
10^{-5}$\,\Msolar\,yr$^{-1}$ (the median mass of those objects is $\sim
0.6$\,\Msolar). The corresponding upper limit is $\sim 6.5$ times this
value, i.e. $\sim 2 \times 10^{-5}$\,\Msolar\,yr$^{-1}$, in excellent
agreement with that found by Cignoni et al. (2009) for this cluster. At
$\sim 20$\,Myr the lower limit to the rate drops by a factor of 30 to
$\sim 10^{-6}$\,\Msolar\,yr$^{-1}$ (median mass $\sim 1$\,\Msolar),
although the corresponding upper limit is about an order of magnitude
higher.


 In spite of the uncertainties on the age of the individual objects
mentioned above, it is clear that beyond $\sim 30$\,Myr a highly
significant increase is seen, both in the star formation rate ($\times
2.5$) and in the number of stars per bin ($\times 5$). This indicates
that the current burst is not the only one in this area and it might not
even be the most prominent one, since the total integrated output of
stars older than $\sim 30$\,Myr and those younger than a few Myr (solid
line) is comparable. In fact, if the fraction of PMS stars with
H$\alpha$ excess has an exponential decay with time (Fedele et al.
2010), episodes occurring more than 30\,Myr ago might have contributed a
large portion of the stars in this region. Since the accuracy on
relative ages is of order a factor of $\sqrt{2}$ (Paper\,I), it is not
possible to estimate exactly how long the previous burst lasted nor how
many short bursts took place in the time frame of $\sim 30$\,Myr covered
by the last age bin. Therefore, if there was just one burst, the star
formation strength might have been considerably higher than that of the
current episode.


Besides very different ages, the two populations of younger and older
PMS stars also have considerably different spatial density
distributions. We compare these distributions to one another in
Figure\,\ref{fig8} by means of contour lines of stellar density with
logarithmic steps, overlaid on a true-colour image of the region. Out of
the total population of 276 PMS stars with well defined masses and ages,
we have selected all those younger than 5\,Myr (a total of 132 objects,
cyan in the online version) and those older than 20\,Myr (in total 111
objects, orange in the online version). The contour plots shown for each
of these groups in Figure\,\ref{fig8} were obtained  after smoothing the
actual distribution with a Gaussian beam with size $\sigma=4\arcsec$ or
$\sim 1.2$\,pc, as indicated by the circle at the bottom of the figure.
The lowest contour level corresponds to a local density of PMS stars
three times as high as the average PMS stars density over the entire
field. The steps between contour levels are constant and corresponds to
a factor of 2. We also show with small dots (yellow in the online
version) the positions of 34 stars with ages between 5 and 20\,Myr.

The remarkable feature in this figure is the difference in the spatial
distribution of younger and older PMS stars: older objects are much more
widely distributed and, except for the centre of NGC\,602, they do  not
always overlap with the younger generation. This difference confirms
that episodic accretion (e.g. Baraffe et al. 2009, 2010) is not at the
origin of the bimodal distribution of PMS stars in Figure\,\ref{fig6},
since it would otherwise require a rather contrived spatial
separation of stars with different accretion histories.  In the
following sections we discuss the two groups of objects in more detail.

\begin{center}
\begin{deluxetable}{ccccccccc}
\tablecaption{Median properties of the stars inside regions 1 -- 4.
\label{tab1}}
\tablehead{
\colhead{ID} & \colhead{IDG} & \colhead{X} & \colhead{Y} &
\colhead{$N_{\rm PMS}$} & \colhead{$m$} & \colhead{$t$} &
\colhead{$L(H\alpha)$} & \colhead{$L^*$}\\
\colhead{} & \colhead{} & \colhead{} & \colhead{} & \colhead{} & 
\colhead{[\Msolar]} & \colhead{[Myr]} & \colhead{[L$_\odot$]} &
\colhead{[L$_\odot$]}}
\startdata
1 & 1 & $-7\arcsec$  & $-20\arcsec$ & 35 & $0.68$ & $1.7$ & $0.003$ & $2.8$ \\
2 & 2 & $-18\arcsec$ & $+15\arcsec$ & 11 & $0.45$ & $0.9$ & $0.003$ & $0.8$ \\
3 &   & $-22\arcsec$ & $-33\arcsec$ & 10 & $0.56$ & $0.1$ & $0.027$ & $1.0$ \\
4 & 4 & $-58\arcsec$ & $-8\arcsec$  &  6 & $0.66$ & $0.6$ & $0.014$ & $1.8$ 
\enddata
\tablecomments{Table columns are as follows: ID number of the region; ID
number as given by Gouliermis et al. (2012); X and Y coordinates in 
Figure\,\ref{fig8}; number of actively accreting PMS stars younger than 
5\,Myr in the region; their median mass, median age,    median H$\alpha$
luminosity, and median bolometric luminosity.}
\end{deluxetable}
\end{center}

\subsection{Younger PMS stars}
\label{ypms}

 PMS stars younger than $\sim 5$\,Myr, traced by the cyan contours
in Figure\,\ref{fig8}, are preferentially located near the central
cluster NGC\,602\,A, which also contains the bulk of massive stars.
Furthermore, the younger PMS stars are distributed in a number of small
groups, some of which match the concentrations of unresolved YSOs
recently identified by Carlson et al. (2011) in their panchromatic study
of these regions covering wavelengths from $0.5$ to 24\,$\mu$m. Most of
these  unresolved YSO sources had already been identified by
Gouliermis et al. (2007) using observations in the range from $0.5$ to
8\,$\mu$m. Both studies conclude that star formation is being
progressively triggered from the cluster centre outwards. This is
inferred from the spatial distribution of the youngest YSOs (Stage I and
I/II), which appear to lie farther from the cluster centre than older
PMS stars, particularly along ridges located at the east-southeast and
west of NGC\,602\,A. 

In fact, what these observations actually reveal is an apparent
systematic age difference between the OB and PMS stars near the cluster
centre and the YSOs along the ridges, in the sense that the latter
appear younger (by  $\sim 2$\,Myr, according to Carlson et al. 2011).
There are presently no measurements to tell whether the massive stars at
the centre are responsible for the formation of the YSOs along the
ridges. Therefore, the observations do not offer direct evidence of
triggering, but rather indicate a case of sequential star formation: the
episode at the centre appears to be already terminated (possibly caused
by ``fuel'' exhaustion, as witnessed for instance by the lower level of
the diffuse $H\alpha$ emission there), while the one along the ridges is
still ongoing. 

Since the groups of unresolved YSOs studied by Carlson et al.
(2011) are located in regions of higher nebulosity, along the ridges
surrounding NGC\,602\,A, it is in principle possible that the reported
age difference between them and the stars in NGC\,602\,A be only
apparent and due to a higher extinction towards the ridges.
Interestingly, also in our photometry most of the youngest PMS  stars
(those with an inferred isochronal age younger than  $0.1$\,Myr) are
located in regions of enhanced  nebulosity, coinciding with the
locations of the groups of unresolved YSOs of Carlson et al. (2011).
While we have assumed the same extinction value for all stars in this
region  ($A_V=0.25$), as  mentioned in Section\,2, if the extinction
towards the youngest PMS stars were about 2\,mag higher, their ages
would become fully compatible with those of the other young PMS objects
at the cluster centre. To understand whether this is the case, we can
look at the column density in these regions.

In the surroundings of NGC\,602, an extinction of $A_V=1$\,mag
corresponds to a column density $N(HI)= (7.4 \pm 1.2) \times
10^{21}$\,cm$^{-2}$ (Gordon et al. 2003). The typical H\,I density in
the heads of the gas pillars along the ridges is about $1.5 \times
10^4$\,cm$^{-3}$, corresponding to a column density of $\sim 2 \times
10^{22}$\,cm$^{-2}$ (Panagia \& De Marchi, in  preparation). This would
in turn generate an extinction of order $A_V \simeq 3$ towards the
objects near the pillars, in good agreement with their observed colours
if they are indeed PMS stars with an age similar to that of the objects
in the cluster's centre. Therefore, in principle, our youngest PMS
stars and the unresolved YSOs of Carlson et al. (2011) could be highly
reddened objects with ages of $\sim 2$\,Myr. On the other hand, some of
the reddest PMS stars in the CMD do not coincide with regions of high 
nebulosity, and it is thus likely that they are intrinsically 
younger than the rest. The question, therefore, remains open as to
whether the PMS stars along the ridges are systematically younger than
those at the cluster centre. A spectroscopic investigation of the
presence and intensity of Lithium absoprtion lines in these stars will
be required to clarify the nature and intrinsic age of these objects.

In the meanwhile, we can set constraints on the relative ages of
these young objects from their H$\alpha$ luminosity $L(H\alpha)$. We
list in Table\,\ref{tab1} the values of $L(H\alpha)$ of all PMS stars
younger than 5\,Myr located inside four sub-clusters, corresponding to
local  concentrations of young PMS stars indicated by the numbers 1
through 4 in Figure\,\ref{fig8}. Stars in regions 3 and 4 are typically
redder in the CMD than those in regions 1 and 2, and comparison with
isochrones results in a younger age. An indication that these stars
are indeed intrinsically younger than those in regions 1 and 2 is
provided by their $L(H\alpha)$, which is a factor of 5--10 higher.
Interestingly, while the level of nebulosity is prominent in region 3,
there is very little of it in region 4, further confirming that these
latter stars are intrinsically very red and therefore intrinsically
young. 

\begin{figure}[t] 
\centering
\resizebox{\hsize}{!}{\includegraphics[width=16cm]{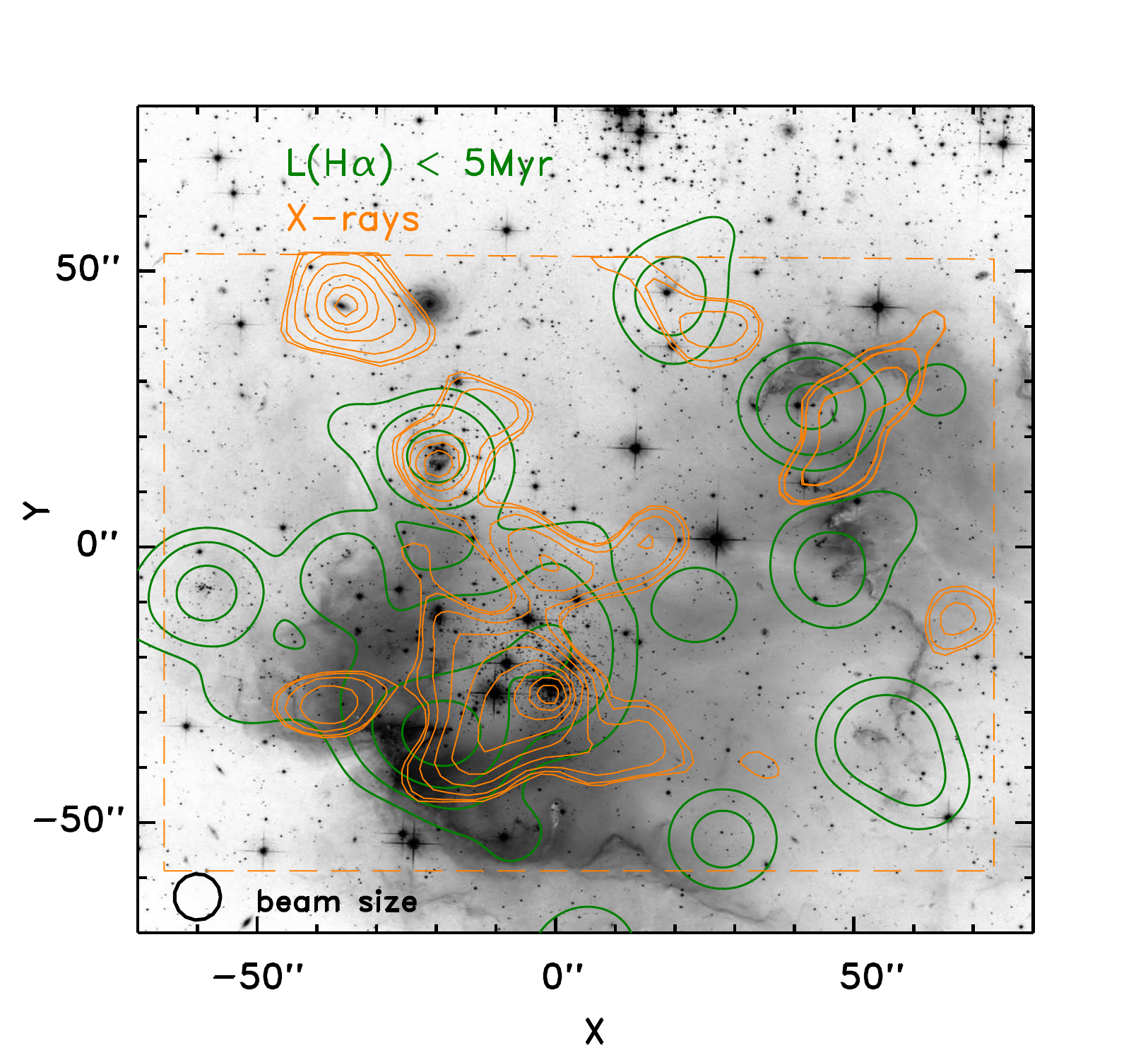}}
\caption{Same as Figure\,\ref{fig8}, but the underlying $H\alpha$ image
is shown in negative, on a grey scale. The cyan contours correspond to
the H$\alpha$ luminosity of PMS stars younger than 5\,Myr, with the
lowest level set to $L(H\alpha)=10^{-5}$\,L$_\odot$ and a constant
logarithmic step of factor of 3. The orange contour lines are taken from
Oskinova et al. (2013) and correspond to levels ranging from $5 \times
10^{-4}$\,cnt\,s$^{-1}$\,arcmin$^{-2}$ to $8 \times 
10^{-4}$\,cnt\,s$^{-1}$\,arcmin$^{-2}$ on a square-root scale. The
dashed  lines define the {\em Chandra} field of view, while the
prominent X-ray source near the top left corner is a background
galaxy.} 
\label{fig9}
\end{figure}

An independent sign of the young ages of the PMS stars identified by the
cyan contour lines in Figure\,\ref{fig8} comes from X-ray observations.
The distribution of our young PMS stars agrees rather well with the 
extended X-ray emission measured by Oskinova et al. (2013) in this field
using {\it Chandra}. For ease of comparison, we show in
Figure\,\ref{fig9} the areas of extended X-ray emission by means of the
contour levels (orange lines) obtained by Oskinova et al. (2013),
together with an intensity map (green lines) of the H$\alpha$ luminosity
from all our PMS stars younger than 5\,Myr. The $H\alpha$ intensity
map differs slightly from the cyan contour lines of Figure\,\ref{fig8},
which represent the number density of PMS stars. The extent of the {\em
Chandra} field of view is marked by the dashed lines. The negative image
in the background is the one obtained in the $H\alpha$ band.

At the distance of NGC\,602, X-ray emission from individual PMS
stars is too faint to be detected, but in  general terms, there is an
overall good match between the peaks of the X-ray and $H\alpha$ maps,
although in the detail some differences are present. For instance, in
the central cluster the peak of the X-ray emission appears to coincide
with the location of the most massive stars, whereas the peak of the
H$\alpha$ emission from low-mass stars is $\sim 10\arcsec$ further
north. This can be understood if, as Oskinova et al. (2013) point out,
the diffuse X-ray emission at the cluster centre originates from (and is
anyhow strongly affected by) a wind-blown bubble. In this case, the
X-ray peak will be determined by the position of the massive objects. 

Another difference concerns region\,4 in Table\,\ref{tab1}, which is
classified by Gouliermis et al. (2012) as a young cluster (also labelled
4 in their list) but that does not reveal diffuse X-ray emission. This
apparent discrepancy, however, stems from the specific detection
limit of the {\em Chandra} observations of Oskinova et al. (2013),
namely $L_X \simeq 10^{32}$\,erg\,s$^{-1}$ (L. Oskinova, priv. comm.),
which prevents clusters of smaller sizes to be detected above the noise.
The study of the Orion Nebula Clusters conducted by Preibisch \&
Feigelson (2005) reveals typical values of $L_X \simeq 3 \times
10^{30}$\,erg\,s$^{-1}$ for  $\sim 1$\,Myr old stars and masses similar
to those in our sample ($0.4 - 1$\,\Msolar). If the same X-ray
luminosity applies to stars in a lower metallicity environment such as
NGC\,602, only clusters with at least 10 -- 15 young sources would be
detectable in the observations of Oskinova et al. (2013). At older ages,
the typical X-ray luminosity decreases (by a factor of $\sim 5$ at an
age of $\sim 20$\,Myr; see Preibisch \& Feigelson 2005), thereby making
it even more difficult to detect these sources. Indeed, none of the
regions in Figure\,\ref{fig8} dominated by PMS stars older than 20\,Myr,
discussed in the following section,  correspond to peaks in the {\em
Chandra} X-ray map.

\begin{figure}[t]
\centering
\resizebox{1.1\hsize}{!}{\includegraphics[bb= 0 0 684 480]{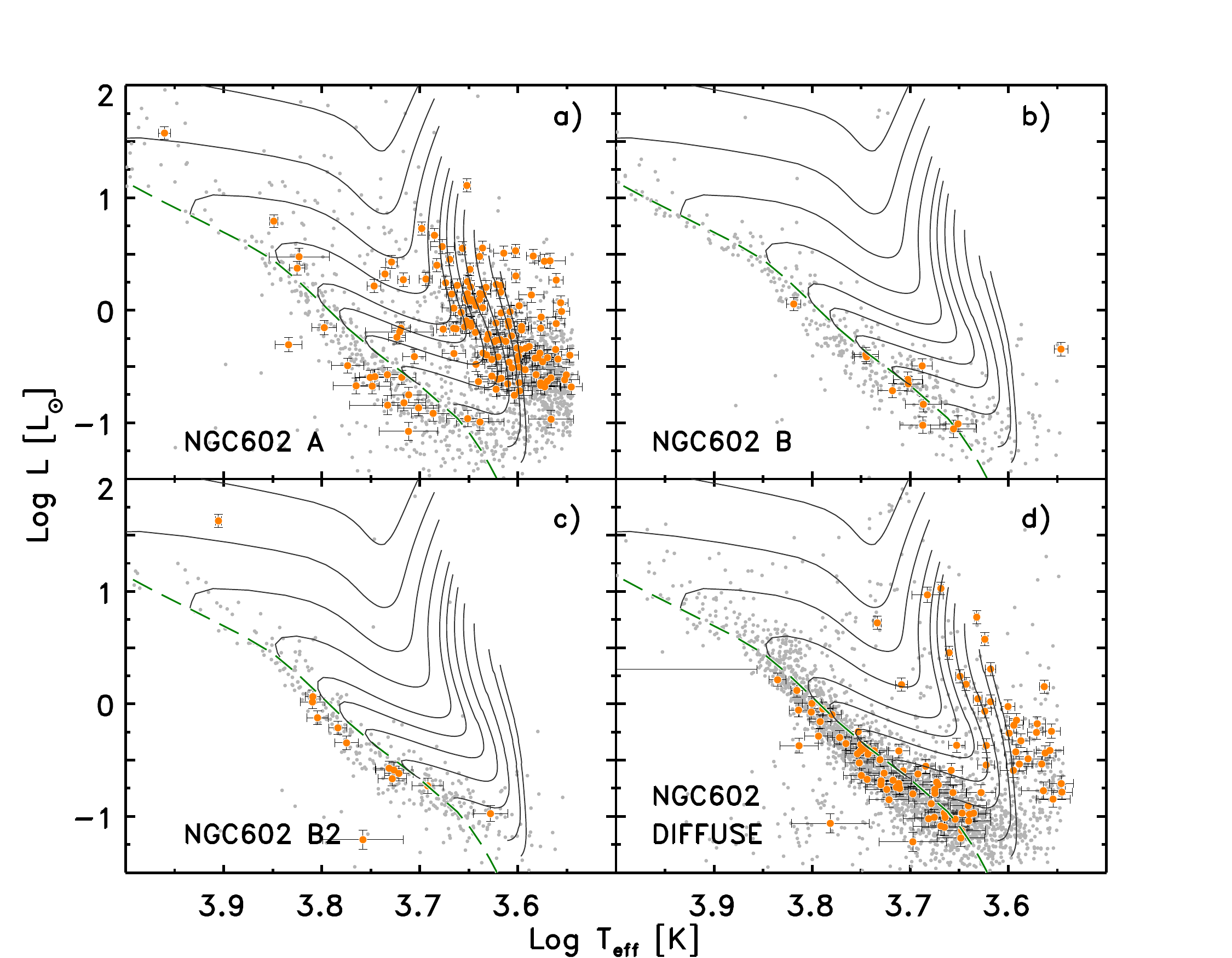}}
\caption{H--R diagrams for the various clusters in the field of
NGC\,602. Objects shown as thick dark dots (orange in the online
version) are PMS stars as in previous figures (error bars on effective
temperature and bolometric luminosity are also shown). Panel a) contains
all stars within $\sim 50\arcsec$ of  the central NGC\,602\,A cluster,
including regions 1--4 in Table\,\ref{tab1}. Panels b) and c) include
all stars within $\sim 25\arcsec$ of the clusters NGC\,602\,B and
NGC\,602\,B2. Panel d) contains the diffuse population in the rest of
the field. Evolutionary tracks are as in Figure\,\ref{fig4}, for masses
of $0.45, 0.5, 0.6, 0.7, 0.8, 0.9, 1.0, 1.2, 1.5, 2.0$, and
$3.0$\,\Msolar from bottom to top.}
\label{fig10}
\end{figure}

\subsection{Older PMS stars}
\label{opms}

Older PMS stars, traced by the orange contours in Figure\,\ref{fig8},
mark the regions where star formation has been active in the recent
past. Some of these objects are seen against regions where gas is still
present, for example the regions of higher nebulosity in the  immediate
surroundings of the central cluster NGC\,602\,A and to the west of it.
The fact that much younger PMS objects are also found in some of these
regions indicates that there is still enough gas to sustain extended
episodes of star formation, although this does not imply any triggering
of star formation but rather a sequential process, in which multiple
generations are born in the same location. In other regions, however,
older PMS stars are not surrounded by younger PMS objects nor by
nebulosity, suggesting that in those places star formation is no longer
active. This is the case of the two small clusters, at the northern edge
of the frame in Figure\,\ref{fig8}, respectively named by Cignoni et al.
(2009) NGC\,602\,B (at $X=+10\arcsec$, $Y=+80\arcsec$) and NGC\,602\,B2
(at $X=+70\arcsec$, $Y=+70\arcsec$). The H--R diagrams of the stars
within $\sim 25\arcsec$ of their centres are compared in
Figure\,\ref{fig10} with those of the central cluster  NGC\,602\,A
(including regions 1--4 in Table\,\ref{tab1}) and  with the diffuse
population in the rest of the field. 

Previous studies had assigned to NGC\,602\,B and NGC\,602\,B2 ages older
than the stars of NGC\,602\,A, albeit with large uncertainties. Schmalzl
et al. (2008) gave upper limits of respectively 180\,Myr and 80\,Myr to
the ages of  NGC\,602\,B and NGC\,602\,B2. These authors also note
that, depending on how the brightest star in NGC\,602\,B is classified,
its age too could be of order  $\sim 80$\,Myr. Cignoni et al. (2009)
compared the CMDs of the two clusters with isochrones based on FRANEC
evolutionary tracks (Chieffi \& Straniero 1989), showing that both are
compatible with ages in the range $15 - 150$\,Myr. The uncertainty,
however, is still very large: these authors acknowledge that the age
differences between these two clusters and the central NGC\,602\,A are
only due to two stars evolved off the MS in NGC\,602\,B and 
NGC\,602\,B2. They conclude that, due to the paucity of stars, the ages
of the two subclusters are practically indistinguishable from that of
NGC\,602\,A.

On the other hand, the presence of a conspicuous number of older PMS
stars still actively accreting, as shown in Figure\,\ref{fig10},
allows us to set much more stringent  constraints on the relative ages
of the two subclusters. We have derived the ages of these PMS objects
using the PMS evolutionary tracks of the Pisa group (Tognelli et al.
2012) using a finer grid of models than the one shown in
Figure\,\ref{fig10}, following the procedure explained in Section\,4.
We find average ages of, respectively, $50 \pm 8$\,Myr for NGC\,602\,B
and $47 \pm 7$\,Myr for NGC\,602\,B2. Based on the quoted uncertainties
and even including possible systematic errors in the PMS evolutionary
tracks, we conclude that the two clusters are coeval and considerably
older than the central cluster NGC\,602\,A. 


A further indication that the two clusters have rather similar ages
comes from the fraction of PMS stars along the MS, i.e. the fraction of
objects that have already reached the MS but that were still accreting
at the time of the observations (see Figure\,\ref{fig10}). For
both clusters we find that this fraction is $\sim 8\,\%$ in the
magnitude range spanned by PMS stars  ($24 \le V \le 27$). Note that
these values are lower limits to the actual fraction of PMS stars,
because they only include objects with H$\alpha$ excess emission at the
$4\,\sigma$ level or more at the time of the observations and because
the objects without excess may also include a number of much older field
stars, not related to these clusters. As mentioned before,  the fraction
of objects with H$\alpha$ excess emission among stars younger than
5\,Myr is about twice as large ($\sim 15\,\%$), but that value too is a
lower limit, since the region of the CMD where these objects fall can be
populated by more massive and older objects subject to differential
reddening. It is reassuring that the ratio of stars with and without
H$\alpha$ excess is larger for younger objects, as expected from stellar
evolution, albeit not as large as the  value of 28\,\% found by De
Marchi et al. (2011b) for the $\sim 1$\,Myr  old PMS stars in NGC\,346.



\section{Mass accretion rate and its evolution}

The physical parameters derived in Section\,4 allow us to determine the
mass accretion rate of the PMS stars. The mass accretion rate $\dot
M_{\rm acc}$ is related to the accretion luminosity $L_{\rm acc}$ via
the free-fall equation, linking the luminosity released by the impact of
the accretion flow with the rate of mass accretion, according to the
relationship:

\begin{equation}
L_{\rm acc} \simeq \frac{G\,M_*\,\dot M_{\rm acc}}{R_*} \left(1 -
\frac{R_*}{R_{\rm in}}\right)
\label{eq1}
\end{equation}

\noindent
where $G$ is the gravitational constant, $M_*$  the mass of the star
determined above, $R_*$ its photospheric radius coming from its
luminosity and effective temperature, and $R_{\rm in}$ the inner radius
of the accretion disc. The value of $R_{\rm in}$ is rather uncertain and
depends on how exactly the accretion disc is coupled with the magnetic
field of the star. Following Gullbring et al. (1998), we adopt $R_{\rm
in} = 5\,R_*$ for all PMS objects and with this assumption we have all
the parameters needed to determine $\dot M_{\rm acc}$. 

\begin{figure}[t]
\centering
\resizebox{\hsize}{!}{\includegraphics[width=16cm]{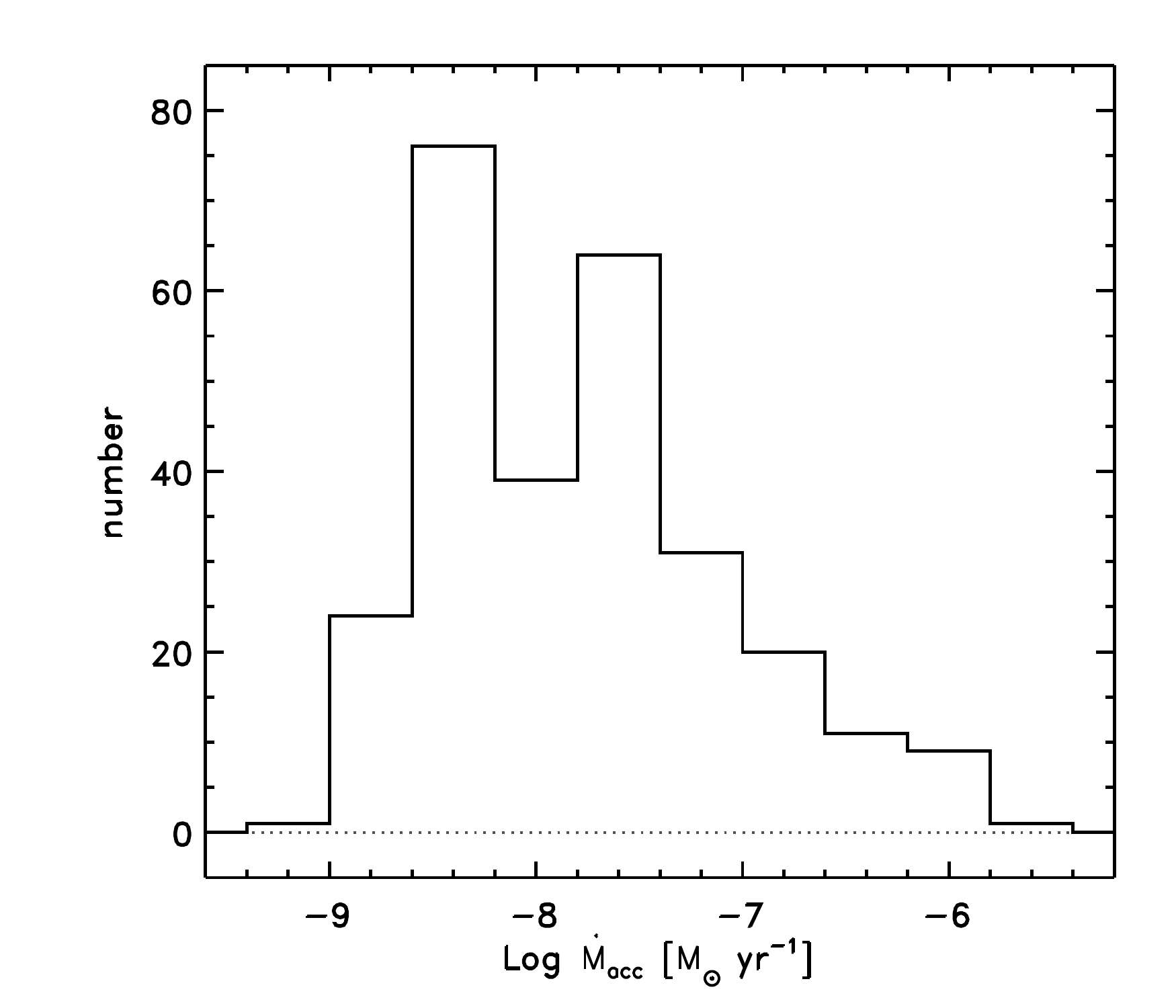}}
\caption{Distribution of the measured mass accretion rates.} 
\label{fig11}
\end{figure}

The median value of $\dot M_{\rm acc}$ that we derive is
$10^{-8}$\,\Msolar\,yr$^{-1}$, but the $\dot M_{\rm acc}$ distribution
shows two clear peaks, at $\sim 4.0 \times 10^{-9}$\,\Msolar\,yr$^{-1}$
and $\sim 2.5 \times 10^{-8}$\,\Msolar\,yr$^{-1}$, suggesting the
presence of two different populations in the field (see
Figure\,\ref{fig11}). The separation between the two peaks is
considerably larger than the observational  uncertainties on $\dot 
M_{\rm acc}$. Papers\,I and II  provide an extensive discussion of the
statistical and systematic uncertainties involved in determining the
mass accretion rate with this method and on its basis we conclude that
the combined statistical uncertainty on $\dot M_{\rm acc}$ for the stars
in our field is 13\,\%. Note that this is just the statistical
uncertainty, but as extensively discussed in Papers\,I and II systematic
uncertainties can dominate, particularly the conversion from
$L(H\alpha)$ to $L_{\rm acc}$. If the uncertainty on the latter relation
were as large as Dahm's (2008) work suggests, it could possibly cause a
systematic effect of up to a factor of 3 on the value of $\dot M_{\rm
acc}$ that we obtain. However, as we explained in Paper II, this
uncertainty cannot be as large as the $0.47$\,dex value suggested by an
elementary fit to the data in the limited compilation of Dahm (2008),
because it has to be intrinsically smaller than the $\sim 0.25$ \,dex
dispersion that we observe in NGC\,346 (Paper\,II). This will also be
visible for NGC\,602 itself in Figure\,\ref{fig13}.

Nevertheless, neither statistical nor systematic uncertainties are
able to explain the double peak in Figure\,\ref{fig11}. Its origin,
however, becomes clearer when the measured $\dot M_{\rm acc}$ values
are compared with the positions of the stars in the H--R diagram. In
Figure\,\ref{fig4}, the size of the  symbols corresponding to the PMS
objects was set to be proportional to $\dot M_{\rm acc}$, as per the
legend in the figure. It is immediately apparent that younger objects
have typically higher mass accretion rates than stars already
approaching the MS, as expected from common wisdom about the theory of
disc evolution.

\begin{figure}[t]
\centering
\resizebox{\hsize}{!}{\includegraphics[width=16cm]{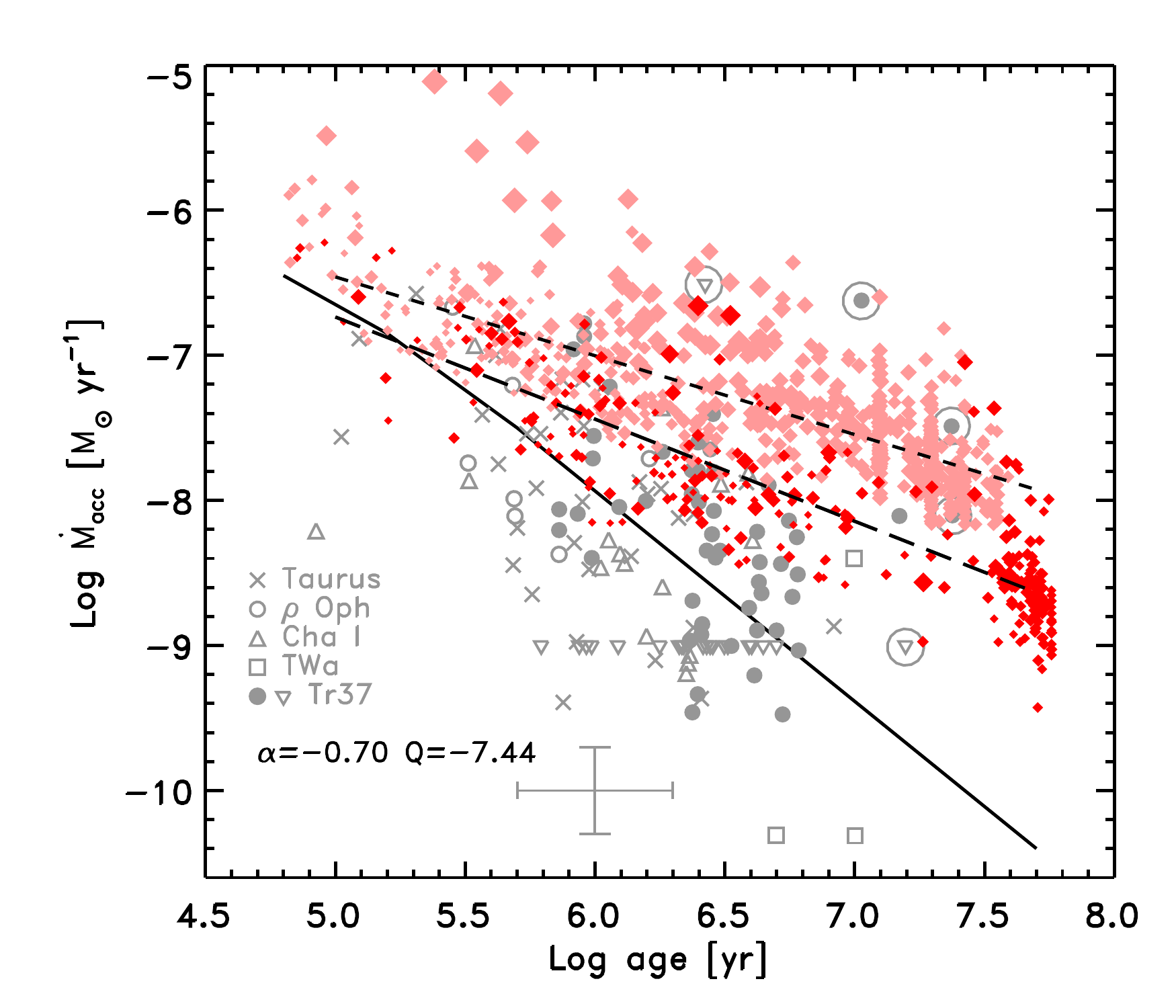}}
\caption{Derived mass accretion rates as a function of the age of the
PMS stars. Dark diamonds (red in the online version) are for stars in
NGC\,602, whereas light grey diamonds (pink in the online version) are 
for the PMS objects in NGC\,346. In both cases the size of the symbols
is proportional to stellar mass. The long- and short-dashed lines are
the best linear fits to the observed distributions in NGC\,602 and
NGC\,346, respectively (the slope $\alpha$ and intercept $Q$ at 1\,Myr
are given for NGC\,602). The other symbols, as per the legend,
correspond to a sample of Galactic  T-Tauri stars studied by
Sicilia--Aguilar et al. (2006), with their typical uncertainty shown by
the large cross at the bottom (the large circles around some of the
symbols indicate dwarfs of spectral type G in the Trumpler 37 sample).
Their distribution, in spite of the large scatter, is consistent with
the models of viscous disc evolution of Hartmann et al. (1998) shown by
the solid line.} 
\label{fig12}
\end{figure}

The wide range of ages and masses covered by the PMS stars in our sample
allows us to study in detail how the mass accretion rate varies with
these parameters. To this end, we show in Figure\,\ref{fig12} the
variation  of $\dot M_{\rm acc}$ as a function of stellar age. The PMS
stars in NGC\,602 are shown as dark diamonds (orange in the online
version), with a symbol size proportional to their mass. Light grey
diamonds in the same figure (pink in the online version) correspond to
the PMS objects identified with the same method in NGC\,346 in
Paper\,II, also with a size proportional to their mass. The other
symbols in the figure (see legend) correspond to nearby T-Tauri stars in
the sample of Sicilia--Aguilar et al. (2006) and are shown here for
reference (the large circles around some of the symbols indicate
dwarfs of spectral type G in the Trumpler 37 sample).

The long dashed line in Figure\,\ref{fig12} represents the best fit to
the observed distribution of mass accretion rates in NGC\,602. Its slope
($\alpha=-0.7$) is rather similar to the one measured in Paper\,II for
stars of similar masses and ages in NGC\,346 ($\alpha=-0.55$), but the
mass accretion rate is systematically lower in NGC\,602 at all masses
and ages. On the other hand, both slopes are considerably less steep
than the $\sim t^{-1.5}$ decline predicted by the models of Hartmann et
al. (1998; see also Calvet et al. 2000; Muzerolle et al. 2000) for
viscous disc evolution, represented here by the solid line, which
reproduces rather well the trend of decreasing $\dot M_{\rm acc}$ with
stellar age for low-mass Galactic T-Tauri stars as compiled by
Sicilia--Aguilar et al. (2006). At face value, this discrepancy would
seem to imply a different evolution of the mass accretion rate for PMS
stars in the Galaxy and in the SMC. However, there are two conceptual
problems with the type of comparison implied by Figure\,\ref{fig12} that
affect the conclusions that one can draw from it: {\em (i)} the masses
of the individual objects must be taken explicitly into account; and
{\em (ii)} the graph offers just a snapshot of the observed 
distribution of mass accretion rates and further assumptions are needed
to infer from it an evolutionary path.

\begin{figure}[t]
\centering
\resizebox{\hsize}{!}{\includegraphics[width=16cm,bb=0 0 530 490]{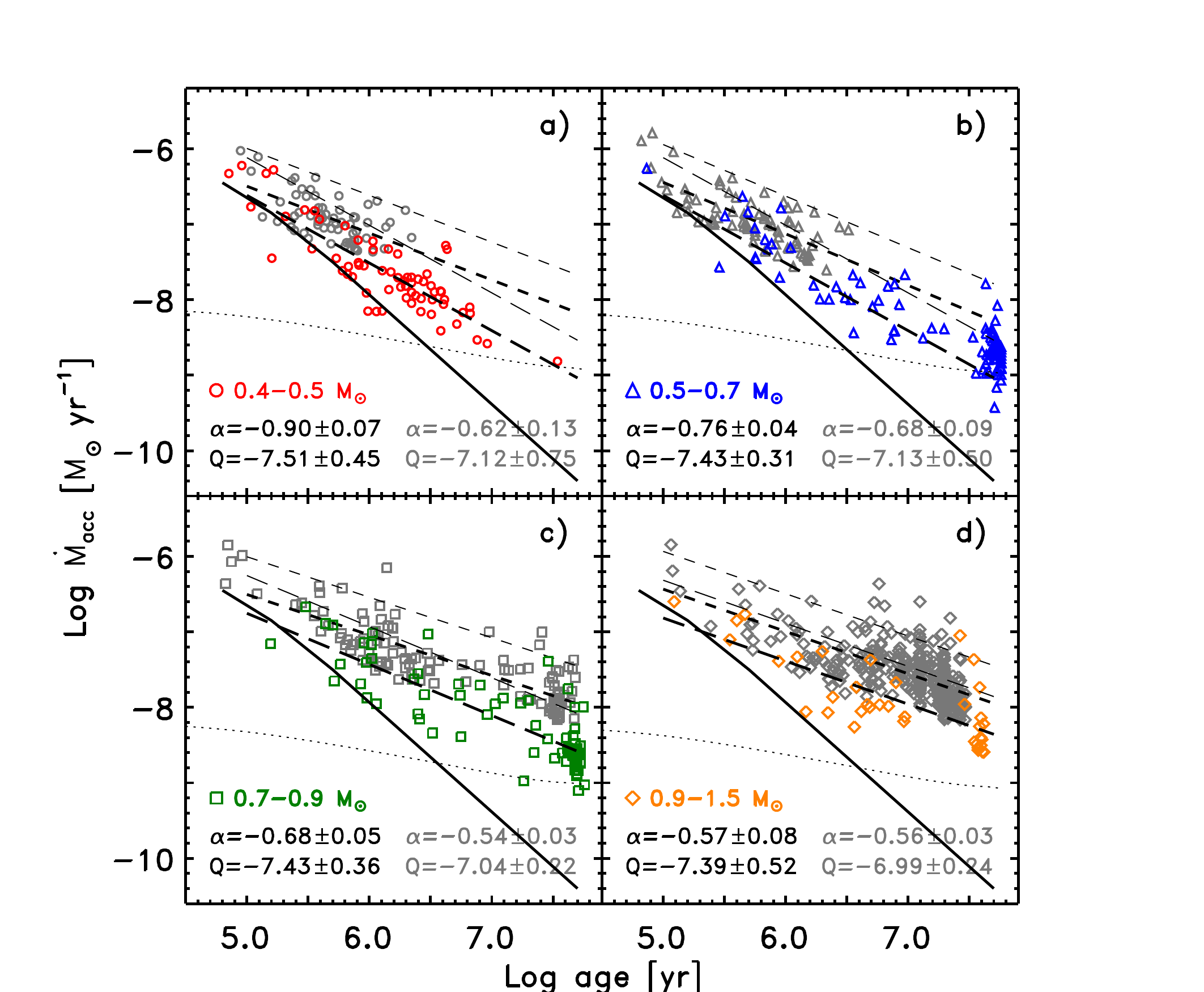}}
\caption{Same as Figure\,\ref{fig12} but for stars in four mass groups,
as indicated in each panel. We also provide the slope $\alpha$ and
intercept $Q$ (at 1\,Myr) of the best fits (thick long-dashed lines).
NGC\,346 data points and fit parameters are indicated in lighter
grey, while the best linear fit is shown by the thick short-dashed
lines. The thick solid line shows the models of Hartmann et al. (1998)
for comparison. The dotted lines correspond to our $L(H\alpha)$
detection limits and are different in each panel owing to the slightly 
different $M/R$ values of each mass group.} 

\label{fig13}
\end{figure}

The first problem is easy to address, since the large size of our sample
allows us to study the mass dependence of $\dot M_{\rm acc}$ in a robust
way. In Papers\,I, II and III we found that the typical $\dot
M_{\rm acc}$ of PMS stars in the LMC and SMC is systematically higher
than for PMS stars of the same mass, independently of their age. To
explore whether that is also the case for NGC\,602, in
Figure\,\ref{fig13} we split our sample in four roughly equally
populated mass groups, namely $0.4 - 0.5$\,\Msolar, $0.5 -
0.7$\,\Msolar, $0.7 - 0.9$\,\Msolar\, and $0.9 - 1.5$\,\Msolar, and plot
for each one separately the run of $\dot M_{\rm acc}$ as a function of
age. The slope $\alpha$ and intercept $Q$ (at 1\,Myr) of the best linear
fit to the data (thick long-dashed lines), according to the relationship
$\log \dot M_{\rm acc} = \alpha \times \log({\rm age}) + Q$ with the age
in Myr, are given in each panel together with their uncertainties. The
same quantities for NGC\,346 are shown in light grey and the
corresponding best fit is indicated by the thick short-dashed lines. As
regards the slopes, the comparison shows that there is good agreement
between the two clusters, although the absence of low-mass stars older
than $\sim 2$\,Myr in NGC\,346 makes the comparison difficult in
Panel\,a). 

A marked difference, however, exists between the slopes of our best fits
and the models of Hartmann et al. (1998; solid lines). As already 
discussed in Papers\,II and III, our stringent requirements on the H$\alpha$
excess emission ($4\,\sigma$) and on the corresponding H$\alpha$
equivalent width (10\,\AA) do in practice set a lower limit to the
H$\alpha$ luminosity that we accept and thus to the $\dot M_{\rm acc}$
value at a given mass or age. On the other hand, comparing the
observations with our $4\,\sigma$ selection thresholds (dotted
lines, independently calculated in  each panel for the average $M/R$
ratio of the stars in that mass group) shows that these effects, if
present, are not important in the determination of the slopes. This is
also confirmed by the fact that the upper envelopes to the observed
distributions of all four mass groups appear to be fully consistent with
the slopes of the best fit. This is shown by the thin long-dashed and
short-dashed lines (respectively for NGC\,602 and NGC\,346), which
represent the  best-fitting line shifted vertically by $0.5$\,dex.

Since our sample is quite rich, we can perform a multivariate
least-square fit to the observations to derive the simultaneous
dependence of $\dot M_{\rm acc}$ on stellar mass and age. We assume a
relationship of the type:  

\begin{equation}
\log \dot M_{\rm acc} = a \times \log t + b \times \log m + c, 
\end{equation}

\noindent
where  $t$ is the age in Myr, $m$ the mass in solar
units and $c$ a constant, corresponding to the intercept at 1\,Myr and
1\,\Msolar\, (note that $c$ is similar to the parameter $Q$ defined above,
but it characterises the simultaneous fit on mass and age). The resulting
best fit gives  $a=-0.72 \pm 0.02$, $b=0.94 \pm 0.14$ and $c=-7.19 \pm
0.24$. The dependence on age and mass are very similar to those found in
Paper\,II for the PMS stars in NGC\,346, namely $a=-0.59 \pm 0.02$,
$b=0.82 \pm 0.09$, confirming the similarity already noticed in
Figure\,\ref{fig13}. As already concluded in Paper\,II for NGC\,346,
using approximate values of $a=-0.6$ and $b=1$ results in a fit with
very small residuals also for NGC\,602. The corresponding values of $c$
are $-7.0$ and $-7.4$ respectively for NGC\,346 and NGC\,602,
reflecting  the fact that the value of $\dot M_{\rm acc}$ is
systematically higher for PMS stars in NGC\,346 than in NGC\,602 at any
given mass or age. This is also visible in Figure\,\ref{fig13}, were for
all mass groups the mass accretion rate in NGC\,346 is consistently
$\sim 2.5$ times higher than in NGC\,602. This difference is meaningful
because both samples are large and cover a wide and overlapping range of
masses and ages. 

As already mentioned, a similar effect had already been noticed in
Papers\,I, II, and III when comparing the mass accretion rates in the
Magellanic Clouds with those of Galactic stars of similar age. It was
suggested in those works that the higher $\dot M_{\rm acc}$ values in
the Magellanic Clouds could be favoured by their lower metallicity. 
As discussed in Paper\,III, the radiation pressure of the forming star
is expected to be less strong on lower-metallicity disc material. Even
though this might not be the dominant effect in the evolution of the
disc, it will delay its dissipation, thereby keeping the accretion
process active for a longer time. In general, a lower metallicity
implies a lower opacity, temperature and viscosity for the disc, and
thus a longer viscous time (e.g. Durisen et al. 2007). Therefore,
low-metallicity stars can undergo significant accretion for a longer
time than higher metallicity stars, in general agreement with our
measurements. In this sense, one could explain the higher observed
$\dot M_{\rm acc}$ values in NGC\,346, since its metallicity is a factor
of two lower than that of NGC\,602, as mentioned in Section\,4. The fact
that the mass accretion rate values in NGC\,602 are about a factor of
$2.5$ lower than those in NGC\,346 at the same mass and age strongly
suggests that the mass accretion rate is a decreasing function of the
metallicity. We will address this dependence in more detail in a
forthcoming paper (De Marchi et al., in preparation), in which we
compare the mass accretion rate values in NGC\,346 and NGC\,602 with
those of stars in other massive clusters in the Galaxy (NGC\,3603;
Beccari et al. 2010) and LMC (30\,Dor; De Marchi et al. 2011c;
Paper\,III).

In summary, our analysis indicates that the mass accretion rate scales
with the age as roughly $t^{-0.6}$, with the first power of the mass and
it decreases with increasing metallicity. However, as mentioned above,
the evolution of $\dot M_{\rm acc}$ with time is derived from the
comparison of the mass accretion rates of stars of different ages.
Unless stars of all ages formed in similar conditions, the observed
snapshot of $\dot M_{\rm acc}$ values may not provide direct information
on the temporal evolution of $\dot M_{\rm acc}$ (see also Natta et al.
2006). In particular, differences in the density, angular momentum or
metallicity of the molecular cloud as it evolves could result in
different star formation properties, leading to different time scales for
the accretion process and the dissipation of the circumstellar discs.

Furthermore, since the fraction of stars undergoing active mass
accretion decreases with time (see Paper\,II), an increasingly larger
number of mildly- or non-accreting stars will fall below our detection
threshold (i.e. the dashed lines in Figure\,\ref{fig13}), thereby
resulting in a shallower slope of the $\dot M_{\rm acc}$ {\em vs.} $t$
relationship at older ages for the least massive stars.  On the other
hand, $\dot M_{\rm acc}$ appears to scale similarly  with age for
different masses and in rather different environments such as NGC\,602,
NGC\,346 and the SN\,1987A field (see Figure\,\ref{fig13} and
Paper\,II). This suggests that density and metallicity do not affect the
way in which $\dot M_{\rm acc}$ varies with time, but rather its
instantaneous value. Indeed, the average value of $\dot M_{\rm acc}$ at
1\,Myr at a specific mass, i.e.  the parameter $Q$ in
Figure\,\ref{fig13}, is systematically higher at higher masses and for
lower metallicity. As such, the parameter $Q$ is less affected by the
uncertainties on the actual dependence of $\dot M_{\rm acc}$ on the
stellar age and can thus more effectively be used to study environmental
differences of the mass  accretion rate.

\section{Summary and conclusions}

We have studied the properties of the stellar populations in the field
of NGC\,602 as observed with the ACS camera on board the HST (proposal
nr. 10248, principal investigator A. Nota; see Carlson et al. 2007),
using a self-consistent method that allows us to reliably identify PMS
stars undergoing active mass accretion, regardless of their age. The
method (see Papers I and II) combines broad-band $V$ and $I$ photometry
with narrow-band $H\alpha$ imaging to identify all stars with excess
H$\alpha$ emission and to derive their accretion luminosity $L_{\rm
acc}$ and mass accretion rate $\dot M_{\rm acc}$. This allows us to
study how star formation and the accretion process have proceeded over
time. The main results of this work can be summarised as follows.

\begin{enumerate}

\item
We have performed PSF-fitting photometry on the exposures in the $V$,
$I$ and $H\alpha$ bands, covering a field of $\sim 3\farcm3 \times
3\farcm3$ around the centre of NGC\,602. From our photometry, we have 
selected all objects with a combined mean uncertainty in the $V$ and $I$
bands not exceeding $0.1$\,mag and a well defined magnitude in the
$H\alpha$ band, for a total of 5\,500 sources.  

\item 
The CMD derived in this way reveals a well defined MS, extending from $V
\simeq 16$ to $V\simeq 28$, a subgiants branch, and a rich population of
young PMS stars, brighter and redder than MS objects. This variety of
stellar populations, already discussed in a number of recent papers
(e.g. Carlson et al. 2007; Cignoni et al. 2009), attests to a complex
star formation history in NGC\,602. Assuming a distance
modulus of $18.92 \pm 0.03$ and the canonical reddening value
$E(B-V)=0.08$, we  find an excellent fit to the upper MS with the Padua
isochrones (Fagotto et al. 1994; Marigo et al. 2008) for metallicity
$Z=0.004$. The same set of models applied to NGC\,346, also in the SMC,
indicate a lower metallicity for the latter cluster, namely $Z=0.002$
(Paper II). 

\item 
Using the self-consistent method developed in Papers I and II, we have
searched for stars with H$\alpha$ excess emission above the $4\,\sigma$
level with respect to the reference provided by normal stars in the same
field. In this way we have identified 296 bona-fide PMS stars by
selecting objects with H$\alpha$ equivalent width in excess of 10\,\AA.
Their median H$\alpha$ luminosity is $2 \times 10^{-3}$\,L$_\odot$.

\item
We have derived effective temperatures and luminosities for these
objects by comparing the photometry to the stellar atmosphere models of
Bessell et al. (1998). In turn, the masses and ages for 276 of these 
objects were obtained by comparing their luminosities and effective
temperatures with the PMS evolutionary models of the Pisa group
(Tognelli et al. 2012). We adopted the metallicity $Z = 0.004$, which is
the value indicated by the comparatively young massive stars in the
cluster. The masses of the 276 PMS stars range from $0.4$\,\Msolar to
$2.5$\,\Msolar, with a median of $\sim 0.7$\,\Msolar. As regards the
ages, there is a clearly bimodal distribution, with about {\small 1/2}
of the objects younger than $\sim 5$\,Myr, about {\small 1/3} older than
$\sim 30$\,Myr, and 50 objects at intermediate ages. This bimodality is
clearly visible in the H--R diagram, where the peaks of their
distributions are separated in effective temperature by $\sim 4$ times
the width of the distributions. Since our age resolution is limited to a
factor of $\sim \sqrt{2}$, stars older than $\sim 30$\,Myr may actually
belong to distinct episodes.

\item
Vigorous episodes of mass accretion separated by longer quiescent
phases, like those proposed by Baraffe et al. (2009, 2010), can be
excluded as the origin of this bimodal distribution in the H--R diagram,
because such a mechanism would be unable to explain the remarkably
different spatial distribution of younger and older PMS stars in this
field. 

\item
PMS stars younger than $\sim 5$\,Myr are distributed in groups near the
cluster centre, as well as along the gas ridges surrounding it, partly
overlapping with regions where unresolved groups of YSOs have been
detected in the mid-IR with {\em Spitzer}. The richest amongst these
groups are also detected by {\em Chandra} as local peaks of the diffuse
X-ray emission above the detection limit of $L_X \simeq
10^{32}$\,erg\,s$^{-1}$, although none of our sources is
individually resolved by {\em Chandra}. Conversely, except for a few
objects near the cluster centre, the distribution of PMS stars older
than $\sim 20$\,Myr does not overlap with that of younger PMS objects
and a large fraction of older PMS stars are found in two clusters
located $\sim 100\arcsec$ north of the centre. The absence of YSOs and
diffuse X-ray emission at these locations confirms that these are not of
the same ``generation'' as younger PMS stars and are intrinsically
older. 

\item 
The relative locations of younger and older PMS stars do not imply a
causal effect or triggering of one generation on the other. The
observations only reveal distinct, and as such most likely unrelated
generations of stars in the same field. We conclude that sequential
star formation is present in this field, without any clear signs of
triggering.

\item
We set a lower limit to the star formation rate for the current ($\sim
2$\,Myr) burst of $\sim 3 \times 10^{-5}$\,\Msolar\,yr$^{-1}$, while at
$\sim 20$\,Myr it  drops by a factor of 30 to $\sim
10^{-6}$\,\Msolar\,yr$^{-1}$. Both values are necessarily lower limits,
since they only account for the stars in the range $0.4 -
2.5$\,\Msolar\, that were undergoing active mass accretion at the time
of the observations. Although at face value the current star formation
episode seems stronger, the total integrated output of the two
populations is comparable. Considering the $\sqrt{2}$ uncertainty on
relative ages in our analysis, if the fraction of PMS stars with
H$\alpha$ excess decays exponentially with time, the episode or episodes
occurring more than $\sim 30$\,Myr ago might have been even stronger
than the current one.

\item
We have derived the mass accretion rate $\dot M_{\rm acc}$ of all
bona-fide PMS stars on the basis of their H$\alpha$ luminosity and the
other physical parameters. The median value of $\dot M_{\rm acc}$ is
$10^{-8}$\,\Msolar\,yr$^{-1}$, but the distribution is clearly bimodal,
with median $\dot M_{\rm acc}$ values of $2.5 \times
10^{-8}$\,\Msolar\,yr$^{-1}$ and $4.0 \times
10^{-9}$\,\Msolar\,yr$^{-1}$ respectively for PMS stars younger than
5\,Myr and older than 20\,Myr. These values are about a factor of $2.5$
lower than those measured in NGC\,346 for PMS stars of similar age.

\item
Our rich sample of PMS stars and its spread in mass and age allow us to
study the evolution of the mass accretion rate as a function of stellar
parameters. Like in the case of NGC\,346 (see Paper II), a multivariate
linear regression fit shows that $\log \dot M_{\rm acc} \simeq
-0.6\,\log t + \log m + c$, where $t$ is the age of the star, $m$ its
mass and $c$ the mass accretion rate for unit mass at 1\,Myr. The value
of $c$ is $-7.4$ for NGC\,602 and $-7.0$ for NGC\,346. The fact that the
mass accretion rate values in NGC\,602 are about a factor of $2.5$ lower
than those in NGC\,346 at the same mass and age suggests  that the mass
accretion may be a decreasing function of the metallicity, as discussed
in Paper\,III.

\end{enumerate}

\begin{acknowledgements}

We are indebted to an anonymous referee, whose constructive criticism
has helped us to improve the presentation of this work. GDM thanks Lidia
Oskinova for useful discussions and Marco Meneghetti for his assistance
with the colour figures. GB aknowledges funding from the European
Community's Seventh Framework Programme (FP7/2007--2013) under grant
agreement No 229517. NP acknowledges partial support by HST-NASA grants
GO-11547.06A and GO-11653.12A, and STScI-DDRF grant D0001.82435. 

\end{acknowledgements}

\end{document}